\documentclass{jpp}
\usepackage{graphicx}
\usepackage{hyperref}
\usepackage[export]{adjustbox}
\usepackage{caption}
 
\usepackage[utf8]{inputenc}
\usepackage[T1]{fontenc}
\usepackage{amsmath}
\usepackage{fancyhdr}
\usepackage{float}
 
\usepackage{orcidlink}
 
\usepackage{comment}
\usepackage{appendix}

\usepackage{subfigure}

\newcommand{\sign}{\text{sign}}

\newcommand{\bm}{\boldsymbol}

\title{Plasma wakes driven by Compton scattering \\ 
Non-linear regime and particle acceleration}

\author{T. Grismayer\aff{1}
  \corresp{\email{thomas.grismayer@tecnico.ulisboa.pt}},
  F. Del Gaudio\aff{1}
 \and L. O. Silva\aff{1}}

\affiliation{\aff{1}GoLP/Instituto de Plasmas e Fus\~ao Nuclear, Instituto Superior T\'ecnico, Universidade de Lisboa, 1049-001 Lisbon, Portugal}

\begin{document}

\maketitle

\begin{abstract}
We investigate plasma wake generation via Compton scattering from photon bursts, a non-ponderomotive process relevant when the photon wavelength is smaller than the interparticle distance but larger than the Compton wavelength. In this regime, electrons can reach relativistic velocities. We extend linear theory to the nonlinear regime, showing that plasma waves can reach the wave-breaking limit. Perfectly collimated drivers produce wakes propagating at the speed of light, allowing electron phase-locking (limited by driver depletion). Non-collimated drivers induce subluminal phase velocities, limiting acceleration via dephasing. Two-dimensional simulations reveal unique transverse fields compared to laser wakefields, with a DC magnetic field leading to consistent focusing. The work considers observational prospects in laboratory and astrophysical scenarios such as around highly luminous compact objects (e.g., pulsars, gamma-ray bursts) interacting with tenuous interstellar or intergalactic plasmas, where conditions favor Compton-dominated wakefield acceleration.
\end{abstract}

\section{Introduction}
Various drivers can generate plasma wakes \cite{Tajima_PRL_1979, Chen1985, Shukla1998}. The multiplicity of potential drivers arises because different types of beams — whether collections of particles or electromagnetic fields — can interact with a plasma through an effective ponderomotive-type force. While the ponderomotive force is most commonly invoked for electromagnetic fields, the concept generalizes to any field or particle ensemble with energy density $\mathcal{E}$ if the driver exerts a force  $F_{\mathrm{pond}} \propto -\nabla \mathcal{E}$. For example, in the case of an optical laser \cite{Tajima_PRL_1979, Esarey1996, Esarey2009} or an X-ray laser \cite{Zhang2016, Svedung2018} $\mathcal{E} \propto E^2 $ with $E$ the laser electric field. For a particle beam driver, $\mathcal{E} \propto n_b$ where $n_b$ is the number density of an electron beam \cite{Chen1985}, and of a neutrino beam \cite{Silva1999}.

The phase velocity of the plasma wake matches that of the driving disturbance and can approach the speed of light. When the wakefield amplitude is sufficiently large, background plasma electrons may become trapped and accelerated to relativistic energies. \cite{Tajima_PRL_1979} first identified this mechanism as a potential alternative to radio-frequency cavities in conventional accelerators and proposed that plasma wakefields could also contribute to the acceleration of cosmic rays in astrophysical environments. Although intense coherent laser pulses are unlikely to occur naturally in astrophysical scenarios, quasi-coherent electromagnetic drivers can be produced in space. Examples that have been proposed include Alfvén shocks \cite{Chen_PRL_2002}, relativistic shocks in which the synchrotron maser instability generates upstream precursor waves \cite{Hoshino_APJ_2008, Kuramitsu_APJL_2008}, fast radio bursts \cite{Petroff2019}, and monster shocks \cite{Beloborodov_2023}.

Recently, \cite{DelGaudio_PRL_2020} demonstrated that a non-ponderomotive mechanism can also drive plasma wakes. Electromagnetic drivers exert two distinct forces on plasma electrons: the familiar ponderomotive force and a second, here termed the Compton force, associated with classical radiation-reaction effects \cite{LandauCTF}. For a monochromatic, collimated light beam, the Compton force is proportional to the energy density ${\bf F}_c = \sigma_T\mathcal{E}$ \cite{peyraud, DelGaudio_PRL_2020}, where $\sigma_T$ is the Thomson cross section. The Compton force exceeds the ponderomotive force for small wavelength or very diluted plasmas, namely $\lambda / r_0 \ll 7.7(r_e/r_0)^{1/4}$, where $r_0 = n_p^{-1/3}$ is the inter-particle distance, $r_e$ the classical electron radius, $\lambda$ the wavelength, and $n_p$ the electron density.

The notion of ponderomotive force requires a quasi-classical electromagnetic field, i.e., the number of photons (of momentum $\hbar k$ with $k = 2\pi/\lambda$) in a volume $k^{-3}$ is large compared with unity, i.e., $E/E_s \gg \sqrt{\alpha}(\hbar k/mc)^2$, where $E$ is the electric field, $E_s = m^2c^3/e\hbar \simeq 1.3 \times 10^{18}~\mathrm{V/m}$ is the Schwinger field, and $\alpha = e^2/\hbar c$. One readily notes that when $\hbar k \sim mc$, the field approaches $E_s$; thus low-frequency fields are generally classical, while very high-frequency fields—if sufficiently weak—cannot be treated classically as discussed in \cite{LandauQED}.

If the inter-particle distance $r_0$ is large compared to $k^{-1}$, the dielectric description fails: photons instead Compton-scatter individual electrons in a dilute ionized gas. The electrons of the plasma are knocked by the photon burst of density $n_{\omega}$ at the frequency $\nu_C = (\sigma_Tn_{\omega}c)^{-1}$. Therefore light sources composed of high-frequency photons propagating in tenuous plasma are likely to interact via Compton scattering with the plasma electrons. 

Astrophysical plasmas cover a large span of densities and radiation environments, making them a fertile ground for exploring Compton-driven wakefield acceleration. While the ponderomotive force has been traditionally considered
\cite{Chen_PRL_2002}, the Compton force can dominate in tenuous plasmas or at shorter wavelengths --- conditions potentially met in various astrophysical settings. From the intergalactic medium ($n_p \sim 0.01 , \text{cm}^{-3}$) to pulsar magnetospheres ($n_p \sim 10^{14} \text{cm}^{-3}$), the inter-particle distance often exceeds 10 $\mu m$. Consequently, compact, energetic astrophysical objects can significantly influence their environments through interactions between radiation and plasma, resulting from the large amounts of nonthermal emission they produce. One example of conditions where the Compton force is relevant is associated with the Eddington limit; in the Thomson regime, this underlies the Eddington luminosity argument. As \cite{Blumenthal1974} demonstrated, the classical Eddington limit, derived assuming Thomson scattering, is modified at higher photon energies due to the Klein-Nishina cross-section. This can significantly alter the radiation pressure exerted on plasma in the vicinity of compact objects. Outflows from accretion disk systems, such as active galactic nuclei and binary systems, may be launched through direct radiation pressure or the Compton rocket \cite{Odell1981, Phinney1982, Henri1991}. Radiative interactions can also lead to drag and deceleration, depending on the specific conditions \cite{Li1992}. \cite{Madau_2000} further explored the radiative acceleration of plasmas in the Klein-Nishina regime, relevant to compact gamma-ray sources. Their work showed that in this regime, particles can be accelerated to asymptotic Lorentz factors at infinity much more rapidly than predicted by Thomson scattering, with a reduced radiation drag due to the relativistic effects on the scattering cross-section. The analysis by \cite{Madau_2000} also highlights the potential for "Compton afterburn" – where random energy imparted to the plasma by gamma-rays is converted to bulk motion, supplementing the direct radiative force. Recent particle-in-cell (PIC) simulations by \cite{Faure2024} delve into the kinetic details of these interactions, revealing a more complex picture. They demonstrate that Compton scattering initially accelerates electrons, but the resulting charge separation also drives ions to relativistic speeds and accelerates other electrons backwards, beyond the driving photon energies. This intricate interplay, coupled with Weibel-type instabilities also shown in \cite{Martinez} and Fermi-like scattering, leads to forward-directed suprathermal electron tails, offering a refined understanding of particle acceleration in extreme radiation environments. The variety of astrophysical sources emitting energetic radiation, such as gamma-ray bursts (GRBs) or pulsars, makes it plausible that Compton-dominated wakefield acceleration is a common and astrophysically relevant process. The implications for particle acceleration in these environments warrant further investigation.

In this article, we extend the one-dimensional linear theory developed in \cite{DelGaudio_PRL_2020} to the nonlinear regime. More specifically, we investigate theoretically and numerically the wake properties and the acceleration of electrons. The analytical results are systematically compared with the particle-in-cell code OSIRIS \cite{OSIRIS}, where a Compton scattering module has been implemented \cite{DelGaudio_JPP_2020}. 
This paper is organized as follows. In section~\ref{sec:linear}, we review the linear regime derived in a previous publication \cite{DelGaudio_PRL_2020}, considering the cases of non-symmetrical photon drivers. Section~\ref{sec:nonlinear} presents the one-dimensional analysis of the nonlinear regime. The acceleration of electrons in the nonlinear wake constitutes the focus of section~\ref{sec:acceleration}. The results obtained in these sections are compared to two-dimensional simulations in section~\ref{sec:2d}. Finally, in section~\ref{sec:conclusions} we state the conclusion and give more concrete numbers/examples about the astrophysical environments that are susceptible to match the conditions for Compton nonlinear wakes to occur.

\section{Linear regime}
\label{sec:linear}
The laser wakefield accelerator (LWFA) concept developed by \cite{Tajima_PRL_1979} relies on a single short high-intensity laser pulse that drives a plasma wake. The wake is driven efficiently when the laser pulse length is on the order of the plasma wavelength $L \simeq \lambda_p$, where $\lambda_p = 2\pi c/\omega_p$. It is reasonable to admit that symmetric and matched coherent or laser drivers can be fine-tuned in the laboratory, but are unlikely to occur in astrophysics. In astrophysical environments, an extremely wide range of frequencies can be generated. In this Section, we review and expand on the linear theory developed in \cite{DelGaudio_PRL_2020}, where the driver is composed of photons only interacting with the electron via Compton Scattering, and extend it considering asymmetric and non-resonant drivers.
 
\begin{figure}
\includegraphics[width=0.8\linewidth, center]{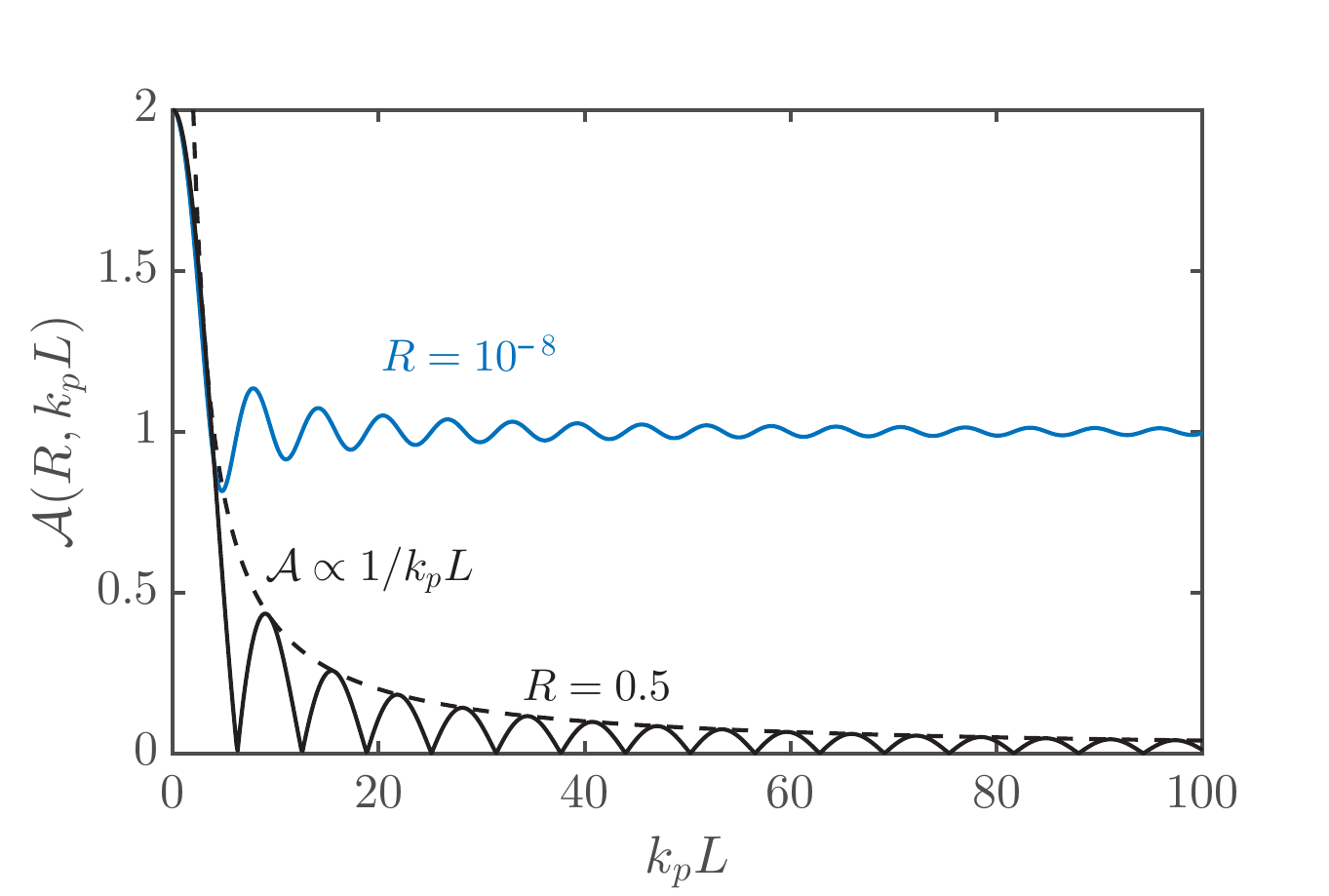}
\caption{Amplitude of the function $\mathcal{A}$ as a function of $k_pL$, given by Eq.(\ref{eq: ampltrig}) plotted in solid line for $R = 10^{-8}$ in blue and for $R=0.5$ in black. For a long driver $k_pL\gg1$, if $Rk_pL\lesssim 1$, the wake amplitude does not decrease. For a symmetric driver $R=0.5$, the wake amplitude falls as $\mathcal{A}\propto 1/k_pL$, shown in dashed line.}
\label{fig: ARL}
\end{figure}

The linear theory previously developed by \cite{DelGaudio_PRL_2020} determines the wake amplitude for a symmetrical photon burst if the energy density $\mathcal{E}$ remains small compared to $\mathcal{E}\ll eE_0/\sigma_T$, where $E_0=mc\omega_p/e$ is the classical cold wave breaking field for a wake with phase velocity close to the speed of light \cite{Dawson1959}. The amplitude of the wake can be computed using the electron fluid equations, continuity for the density $n$, momentum for the fluid velocity $\beta c$, and Poisson's equation for the longitudinal field $E_x$. The averaged fluid force exerted by the photon burst is $\sigma_T\mathcal{E}$.
The set of equations reads
\begin{eqnarray}
\frac{\partial n}{\partial t} &=& -c\frac{\partial}{\partial x} \left(n\beta\right)~, \\
mc\frac{d \beta}{dt} &=& -eE_x + \sigma_T\mathcal{E}~. \\
\frac{\partial E_x}{\partial x}  &=& 4\pi e(n_p-n)~,
\end{eqnarray}
where $n_p$ is the plasma density. The ions are assumed to be immobile. The linearized version of this set of equations is
\begin{eqnarray}
\frac{\partial n_1}{\partial t} &=& -cn_p\frac{\partial \beta_1}{\partial x}~,  \\
mc \frac{\partial \beta_1}{\partial t} &=&  -e E_x + \sigma_T\mathcal{E}~, \\
\frac{\partial E_x}{\partial x}  &=& -4\pi e  n_1~,
\end{eqnarray}
where the electron density is written as $n=n_p+n_1$. The plasma is considered to be initially at rest, implying $\beta=\beta_1$.
The change of variables $\tau=\omega_p t$, $\xi = k_p(x-ct)=k_px-\tau$, $k_p=\omega_p/c$, leads to
\begin{eqnarray}
\frac{\partial}{\partial\tau}\left(\frac{n_1}{n_p}\right)&=& \frac{\partial}{\partial\xi}\left(\frac{n_1}{n_p}-\beta_1\right)~,\\
\frac{\partial\beta_1}{\partial\tau}&=&\frac{\partial\beta_1}{\partial\xi}-\frac{E_x}{E_0}+\frac{\sigma_T\mathcal{E}}{mc\omega_p}~,\\
\frac{\partial }{\partial \xi}\left(\frac{E_x}{E_0}\right)  &=& -\frac{n_1}{n_p}~.
\end{eqnarray}
The quasi-static approximation, commonly used in the theory of laser wakefield acceleration \cite{Sprangle_1990, Birdsall}, neglects $\partial / \partial \tau$ in the electron-fluid equations, which is valid when the driver evolution time is long compared to the transit time of the plasma through the driver, i.e, $\omega \gg \omega_p$ for a laser of frequency $\omega$. In the case of an uncoherent photon driver (made of photons of frequency $\omega$ and density $n_{\omega}$), the typical evolution time of the burst envelope is $t_{\omega} = \nu_{\omega}^{-1} = (\sigma_Tn_pc)^{-1}$. The quasi-static approximation in this case amounts to $\nu_{\omega} \ll \omega_p$, which is equivalent to $n_p \ll n_{\omega}$ (provided that $\nu_C / \omega_p \gtrsim 1 $). The equation for the electrostatic field is now
\begin{equation} \label{eq: linear}
\left(\frac{\partial^2}{\partial\xi^2}+1\right) \frac{E_x}{E_0} = \frac{\sigma_T\mathcal{E}}{mc\omega_p}.
\end{equation}
The solution for the longitudinal field is given by the convolution of the source term with the Green function of the harmonic operator
\begin{equation}
\label{eq:conv}
\frac{E_x}{E_0} =\frac{\sigma_T\mathcal{E}_0}{mc\omega_p} \int_0^{-k_pL} \sin(\xi-s) f(s)ds,
\end{equation}
where $\mathcal{E}(s)=\mathcal{E}_0f(s)$, $f$ being the bounded shape function of the driver, and $L$ the length of the driver.
The integration by parts of the integral $I=\int_0^{-k_pL}~ds \sin(\xi-s) f(s)$ yields
\begin{equation} \label{eq: integralI}
I = \cos(\xi-s)f(s)\Big\lvert_0^{-k_pL}+\int_0^{-k_pL}~ds \cos(\xi-s) \frac{\partial f}{\partial s}~,
\end{equation}
where $f(s)\Big\lvert_0^{-k_pL}=0$ is imposed since $f$ is bounded. 
To understand the influence of the shape of the driver on the amplitude of the wake, we consider a driver of triangular shape characterized by a rise time $RL/c$ and a fall length $(1-R)L/c$, where $R$ is the rise time normalized to $L/c$.
With the choice of $f$, the integral given by Eq.(\ref{eq: integralI}) has an exact solution
\begin{eqnarray} 
I &=& \frac{1}{k_p}\mathcal{A}\cos(\xi+\theta), \\
\mathcal{A}&=&\sqrt{a^2+b^2}~, \label{eq: ampltrig}\\
\theta&=&\arctan(b/a)~.
\end{eqnarray}
where
\begin{eqnarray}
a&=& \frac{\cos(Rk_pL)-1}{Rk_pL}+\frac{\cos(k_pL)-\cos(Rk_pL)}{(1-R)k_pL}~, \label{eq: sinf}\\
b&=& \frac{\sin(Rk_pL)}{Rk_pL}+\frac{\sin(k_pL)-\sin(Rk_pL)}{(1-R)k_pL}~. \label{eq: cosf}
\end{eqnarray}
From  Eq.(\ref{eq: sinf}) and Eq.(\ref{eq: cosf}), we observe that whenever $Rk_pL\lesssim 1$ or $(1-R)k_pL\lesssim1$, the wake amplitude $\mathcal{A}\sim 1$ and does not diminish with respect to $k_pL$, even for non-resonant long drivers $k_pL\gg 1$.
If instead $Rk_pL\gg1$ or $(1-R)k_pL\gg1$ then $\mathcal{A}\propto 1/k_pL$.

\begin{figure}
\includegraphics[width=0.8\linewidth, center]{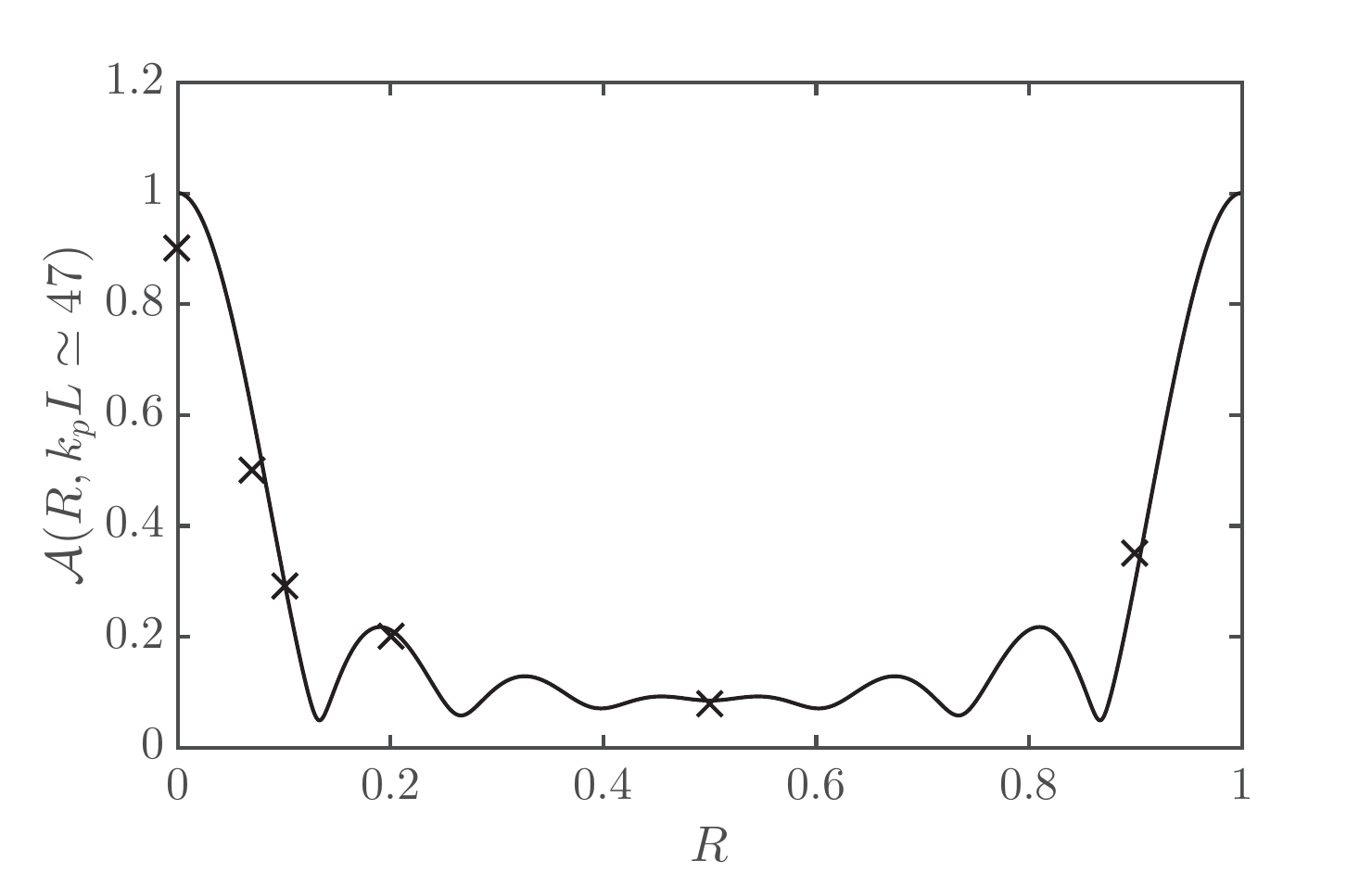}
\captionsetup{justification=raggedright}
\caption{Amplitude of the function $\mathcal{A}$ as a function of $R$, given by Eq.\ref{eq: ampltrig} plotted in solid line for $k_pL\simeq47$. 1D PIC simulation results are displayed with ($\times$).}
\label{fig: A_R_nRel}
\end{figure}

These scalings are shown in Fig. \ref{fig: ARL} where the amplitude of the function $\mathcal{A}$ is plotted as a function of $k_pL$ for two cases $Rk_pL\ll1$ and $R=0.5$.
We have also verified these scalings with 1D particle-in-cell simulations. The burst is composed of photons of energy $\hbar\omega=0.001~mc^2$ with $\mathcal{E}_0=0.01eE_0/\sigma_T$ propagating in a cold plasma of uniform density $n_p=1~\mathrm{cm^{-3}}$. The simulations are performed with a moving window $100~d_e$ long ($d_e = k_p^{-1}$), with $\Delta x = 0.01k_p^{-1}$, and $\Delta t = 0.0099~\omega_p^{-1}$. The number of particles in each cell is $100$, and the burst is made of photons of energy $\hbar\omega=0.001~mc^2$ with $\mathcal{E}_0=0.01eE_0/\sigma_T$. Figure \ref{fig: A_R_nRel} shows the amplitude of the function $\mathcal{A}$ as a function of $R$, given by Eq. \ref{eq: ampltrig} for $k_pL\simeq 47$ obtained from Eq. \ref{eq: ampltrig}, in solid line, and from simulations, denoted by $(\times)$, demonstrating the excellent agreement between theory and simulations. 

\section{Nonlinear regime}
\label{sec:nonlinear}
We consider now the case of a photon burst with an energy density sufficiently high to push electrons into relativistic motion. The nonlinear regime of plasma wakes, which has been extensively studied for laser drivers by \cite{Sprangle_1990, Esarey1996}, is reached when the plasma electric field approaches the wavebreaking limit. In our case, the photon energy density must be capable of generating a wake amplitude $E_0$ --- the photon energy density being on the order of $\mathcal{E}_0 \sim eE_0/\sigma_T$. Using the definition of $E_0$, it corresponds to a photon energy density of 
\begin{equation}
\label{eq:Ene_den_nl}
    \mathcal{E}_0 \sim 10^{12}\sqrt{n[\mathrm{cm^{-3}]}} ~\mathrm{erg~cm^{-3}}
\end{equation}
The amplitude of the Langmuir wake driven by the burst can be derived solving the following system of equations
\begin{eqnarray}
\nonumber
\frac{\partial n}{\partial t} &=& -c\frac{\partial}{\partial x} \left(n\beta\right) \\
\nonumber
mc\frac{d \gamma\beta}{dt} &=& -eE_x + \sigma_T\mathcal{E}(1-\beta) \\
\frac{\partial E_x}{\partial x}  &=& 4\pi e(n_p-n) 
\label{NLfluideq}
\end{eqnarray}
where the momentum equation includes the relativistic fluid version of the Compton force (see Appendix \ref{app: CMP_rel_force}). The relativistic Compton force used here is valid when $\mathcal{E}_0\gamma(1-\beta) \ll n_{\omega}mc^2$, which limits the photon energy to $\gamma mc^2$ if $\gamma \gg 1$. A version of the relativistic Compton force valid for all photon energies can be found in \cite{Blumenthal1974} and \cite{Faure2024}. Rewriting the equations (\ref{NLfluideq}) as a function of $\tau$ and $\xi$, and using the quasi-static approximation, we obtain 
\begin{eqnarray}
\nonumber
\frac{\partial}{\partial\xi}\left(n(1-\beta)\right)&=&0\\
\nonumber
\frac{\partial}{\partial\xi}\left(\gamma(\beta-1)\right)&=&\frac{E_x}{E_0}-\frac{\sigma_T\mathcal{E}}{mc\omega_p}(1-\beta) \label{eq: momQS}\\
\frac{\partial }{\partial \xi}\left(\frac{E_x}{E_0}\right)  &=& 1-\frac{n}{n_p}
\label{NLfluid2}
\end{eqnarray}
The continuity equation implies $n(1-\beta)=n_p$, and Poisson's equation in equations (\ref{NLfluid2}) can thus be rewritten as $\partial_{\xi}(E_x/E_0) = \beta/(\beta-1)$. Deriving the momentum equation with respect to $\xi$ and substituting into Poisson's equation, we arrive at a second-order equation for $\beta$
\begin{equation} \label{eq: nonlinear}
\frac{\partial^2}{\partial\xi^2} \left[\gamma(\beta-1) \right]  +\frac{\beta}{1-\beta} + \frac{\partial}{\partial\xi} \left[\frac{\sigma_T\mathcal{E}}{mc^2} (1-\beta)\right]=0.
\end{equation}
In the limit $\beta \ll 1$, the linear theory is recovered. Poisson's equation becomes $\partial_{\xi} (E_x/E_0) = -\beta$, and the momentum equation Eq.(\ref{eq: momQS}) becomes $\partial_{\xi^2} \beta +\beta + \partial_{\xi}\sigma_T\mathcal{E}/mc\omega_p= 0$.

The nonlinear theory for the laser wakefield, reviewed by \cite{Esarey1996}, determines the maximum amplitude of a driven plasma wave. Behind the driver, where $\mathcal{E}=0$, the conditions $\partial_{\xi} (E_x/E_0)=\beta/(\beta-1)$ and $\partial_{\xi}\left[\gamma(\beta-1)\right]=(E_x/E_0)$ hold. Simple algebra shows that these equations are equivalent to $\partial_{\xi}(E_x/E_0)^2/2+\partial_{\xi}\gamma=0$, which can be interpreted as an energy conservation law
\begin{equation}
\frac{1}{2}\frac{E_x^2}{E_0^2} + \gamma = \hat{\gamma}~,
\end{equation}
where the $E_x^2$ term is the energy density stored in the field, $\gamma$ is the normalized energy of the electron fluid, and $\hat{\gamma}$ is a constant given by the maximum Lorentz factor of the electron fluid, when $E_x=0$. 
The field amplitude is maximum when the electron fluid is at rest $\gamma=1$, leading to
\begin{equation} \label{eq: Rel E}
E_{\mathrm{max}} = E_0\sqrt{2(\hat{\gamma}-1)}.
\end{equation}
This is the expression of the cold relativistic wave breaking limit \cite{Esarey1996, Akhiezer} when $\hat{\gamma} = \gamma_{\phi}$, where $\gamma_{\phi}$ is the relativistic factor associated with the phase velocity of the plasma wave. 
As already demonstrated in the previous work by \cite{DelGaudio_PRL_2020}, and further discussed in this article, a collimated photon driver propagates at the vacuum speed of light inside the plasma, which implies $\gamma_{\phi}\rightarrow \infty$. Nonetheless, the value of $\hat{\gamma}$ must be finite for a given finite $\mathcal{E}_0$. To estimate $\hat{\gamma}$ under these assumptions, we consider the balance between the potential energy stored in the electric field at the maximum displacement $\hat{\xi}$ and the work done by the Compton force during that displacement.
\begin{equation} \label{eq: work balance}
\int_0^{\hat{\xi}}\frac{\sigma_T\mathcal{E}}{mc\omega_p}(1-\beta)d\xi\simeq  \frac{E_x}{E_0}\hat{\xi}~.
\end{equation}
Note that Eq. (\ref{eq: work balance}) holds only within the first plasma oscillation, thus for short or resonant drivers $k_pL\lesssim 2\pi$. In the case of a resonant driver with shape $\mathcal{E} =\mathcal{E}_0 \sin^2\left(\xi/2\right)$, $\xi \in [-2\pi,0]$, we approximate
\begin{equation} \label{eq: approx F_c}
\int_0^{\hat{\xi}}d{\xi}\frac{\sigma_T\mathcal{E}}{mc\omega_p}(1-\beta)\simeq \frac{\pi}{2}\frac{\sigma_T\mathcal{E}_0}{mc\omega_p}\left(1-\frac{\hat{\beta}}{2}\right)\hat{\xi}~,
\end{equation}
where $\hat{\beta}/2 \simeq {\hat \xi}^{-1}\int_0^{\hat \xi}\beta d\xi$.
By replacing Eq.(\ref{eq: Rel E}) and Eq.(\ref{eq: approx F_c}) into Eq. (\ref{eq: work balance}) we obtain
\begin{equation}\label{eq: hatgamma}
\sqrt{2(\hat{\gamma}-1)}\simeq \frac{\pi}{4}\frac{\sigma_T\mathcal{E}_0}{mc\omega_p}\frac{2\hat{\gamma}-\sqrt{\hat{\gamma}^2-1}}{\hat{\gamma}}
\end{equation}
which relates the energy density $\mathcal{E}_0$ to the maximum Lorentz factor $\hat{\gamma}$ of the electron fluid.
The nonlinear wakefield amplitude is obtained from Eq. (\ref{eq: Rel E}) where $\hat{\gamma}$ is a solution of Eq.(\ref{eq: hatgamma}). When $\hat{\gamma} \gg 1$, the right-hand side of Eq.(\ref{eq: hatgamma}) tends towards a constant value. Therefore, in the relativistic regime for a resonant driver
\begin{eqnarray} 
\label{eq:Emaxrel}
E_{\mathrm{max}} \simeq \frac{\pi}{4}\frac{\sigma_T\mathcal{E}_0}{e}.
\end{eqnarray}
This indicates that, similar to the linear regime, the amplitude of the wake is directly proportional to the photon energy density. 

\begin{figure}
\includegraphics[width=0.8\linewidth, center]{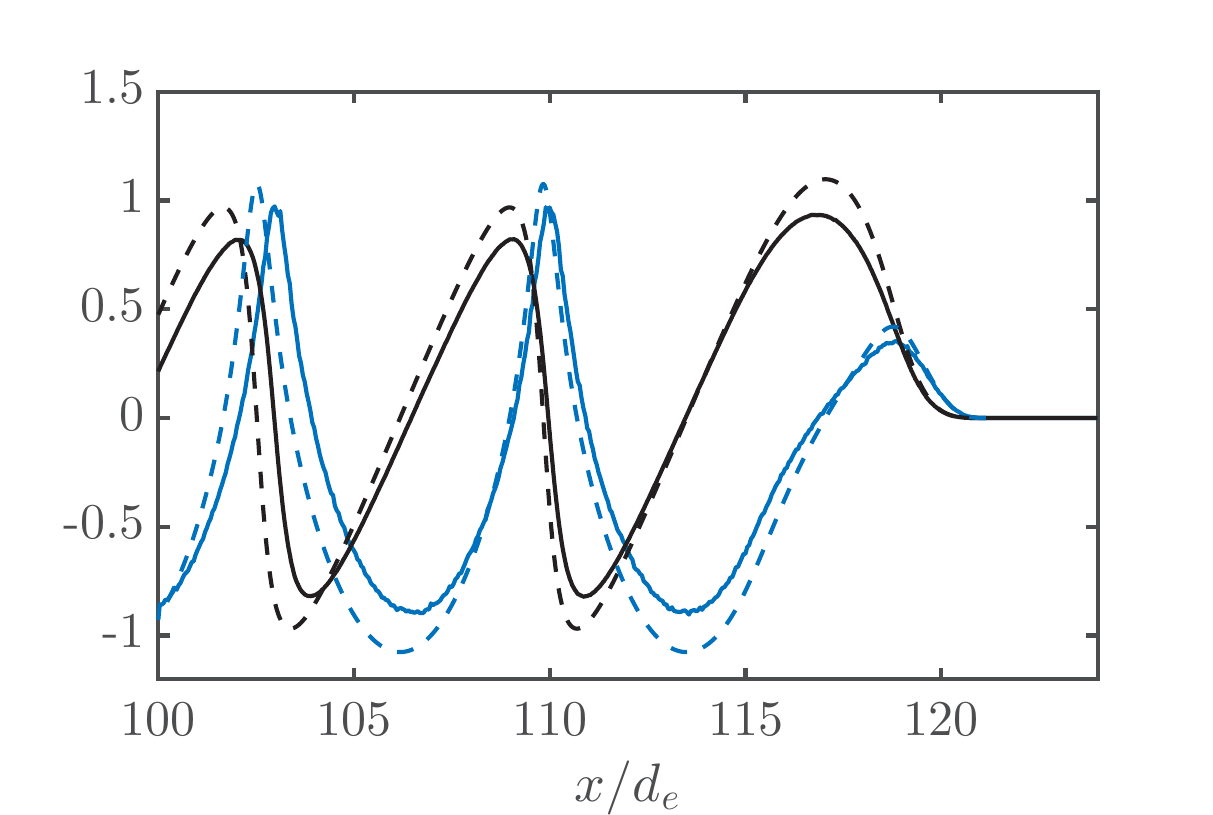}
\captionsetup{justification=raggedright}
\caption{Amplitude of the wake field (black) and normalized fluid momentum of the plasma electrons (blue) for a resonant driver with energy density of $\mathcal{E}_0=eE_0/\sigma_T$. The simulation results are represented with a solid line and the theory, solution of Eq.(\ref{eq: nonlinear}), with a dashed line.}
\label{fig: nlwake}
\end{figure}

\begin{figure}
\includegraphics[width=0.8\linewidth, center]{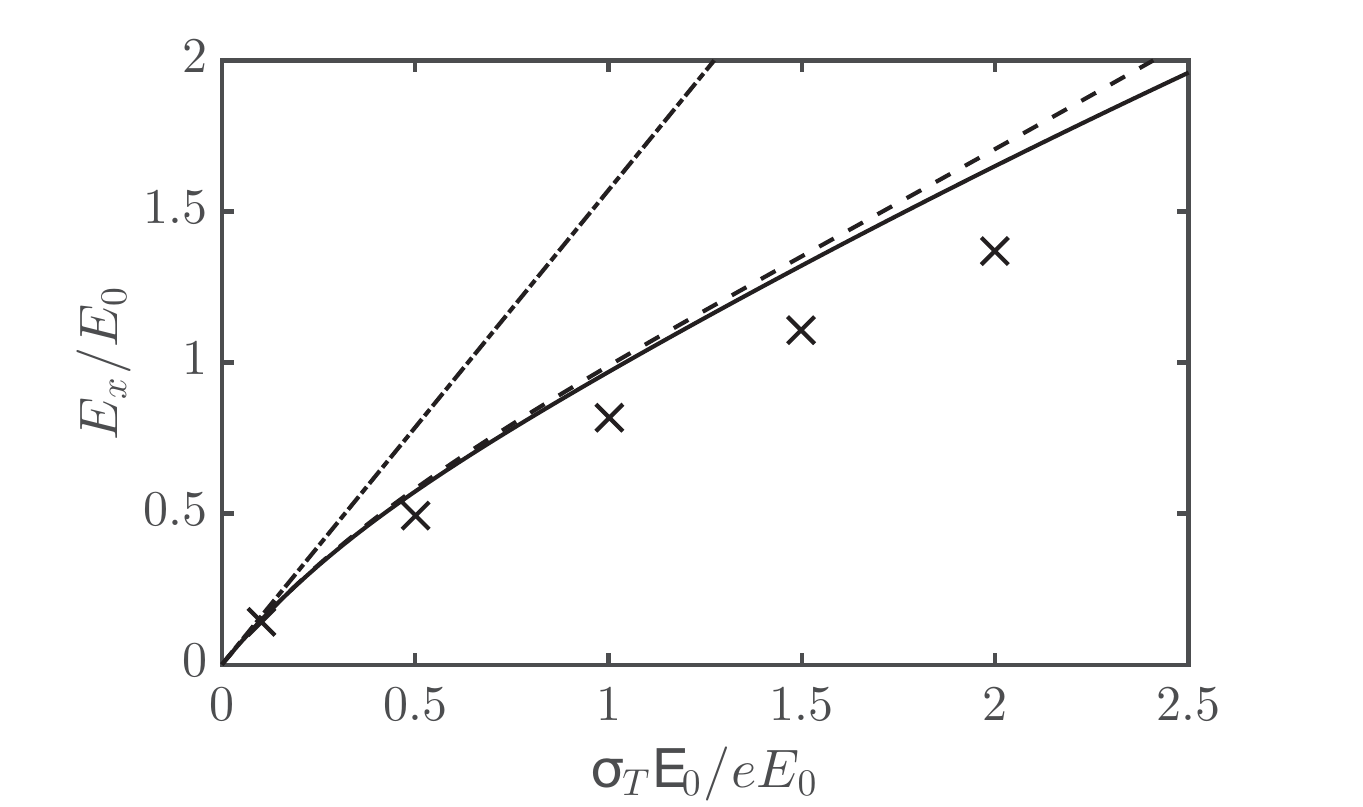}
\captionsetup{justification=raggedright}
\caption{Amplitude of the wakefield as a function of the driver energy density. Simulations are denoted with $(\times)$, the numerical solution of Eq (\ref{eq: nonlinear}) is in solid line, Eq. (\ref{eq: Rel E}) with $\hat{\gamma}$ obtained from Eq. (\ref{eq: hatgamma}) in dashed line, and linear theory is in dot-dashed line.}
\label{fig: nlwake_scaling}
\end{figure}

\begin{figure}
\includegraphics[width=0.8\linewidth, center]{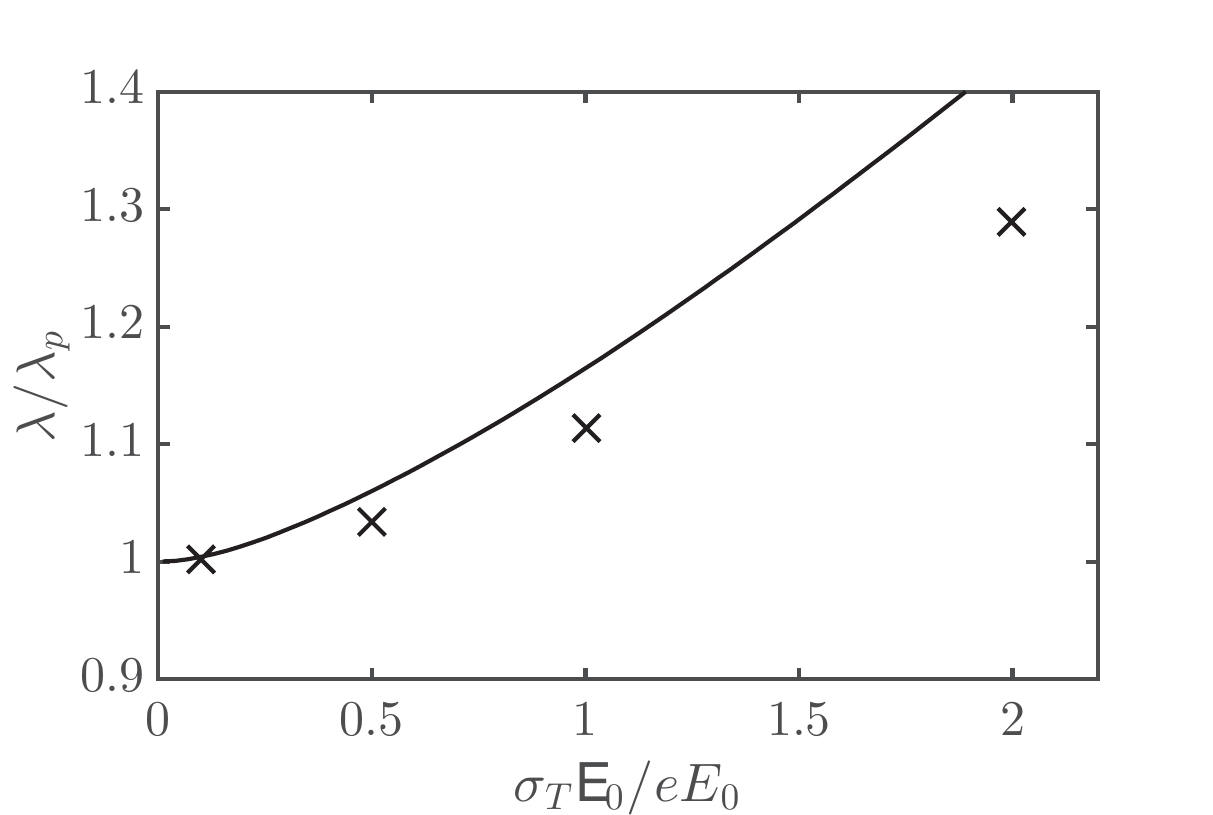}
\captionsetup{justification=raggedright}
\caption{Wavelength of the wake as a function of the driver energy density. Simulations are denoted with $(\times)$, the solid line is the numerical solution of Eq (\ref{eq: nonlinear}).}
\label{fig: scale_lambda}
\end{figure}

Figure \ref{fig: nlwake} shows the amplitude of the wakefield $E_x/E_0$ (in black lines) and the fluid momentum of the plasma electrons $\gamma\beta$ (in blue lines) for a peak energy density $\mathcal{E}_0=eE_0/\sigma_T$. The maximum momentum of the electron fluid over the plasma oscillations is $\gamma\beta\simeq1$, which implies a maximum Lorentz factor of $\hat{\gamma}\simeq1.4$, in agreement with Eq.(\ref{eq: hatgamma}) for $\sigma_T\mathcal{E}_0 = e E_0$. We compared the numerical solution of Eq.(\ref{eq: nonlinear}) and the predictions of Eqs.(\ref{eq: Rel E}) and Eq.(\ref{eq: hatgamma}) with 1D PIC simulations in the nonlinear regime $\sigma_T\mathcal{E}_0\sim eE_0$. The simulations are performed with a moving window in a box $24k_p^{-1}$ long, filled with a cold plasma of density $n_p=1~\mathrm{cm^{-3}}$. The spatial resolution is chosen to be $\Delta x = 0.001k_p^{-1}$, the time step $\Delta t = 0.00099~\omega_p^{-1}$, and the number of particles per cell is $30$.
The burst is composed of photons of energy $\hbar\omega=0.01~mc^2$, with a Gaussian shape $\mathcal{E} =\mathcal{E}_0 \cos^2\left(\pi\xi/L\right)$, $\xi \in [-L,0]$, and is resonant $L=\lambda_p$ (the duration of the burst is $\tau_p=L/c \simeq 0.1\mathrm{ms}$) to maximize the amplitude of the electric field.

Figure \ref{fig: nlwake_scaling} illustrates the amplitude of the wakefield as a function of the peak energy density for a resonant driver. As expected, the results deviate from linear theory, represented by the dot-dashed line. The nonlinear theory, described by Eq. (\ref{eq: nonlinear}) and shown with a solid line, aligns well with particle-in-cell (PIC) simulation results, indicated by $(\times)$. Additionally, the dashed line corresponds to Eq. (\ref{eq: Rel E}) with $\hat{\gamma}$ obtained from Eq. (\ref{eq: hatgamma}), and it agrees with the numerical solution of Eq. (\ref{eq: nonlinear}). At higher energy densities, the theory increasingly diverges from simulation outcomes, which indicate the kinetic effects -- absent in cold fluid models -- become significant. We conjecture that a portion of the driver energy is diverted into plasma thermal energy rather than solely contributing to electron fluid motion. Figure \ref{fig: scale_lambda} depicts the wake’s wavelength as a function of the peak energy density $\mathcal{E}_0 = eE_0/\sigma_T$. The nonlinear theory (Eq. \ref{eq: nonlinear}) (solid line) agrees with PIC simulations (denoted by $(\times)$). As electron velocities approach relativistic speeds, plasma oscillations decrease in frequency ($\omega_p \to \omega_p/\sqrt{\hat \gamma}$) and experience an increase in wavelength ($\lambda_p \to \sqrt{\hat \gamma}\lambda_p$), due to the relativistic increase in electron mass.

\section{Acceleration of leptons}
\label{sec:acceleration}
Plasma wakes with phase velocities approaching the speed of light have been first demonstrated experimentally by \cite{Clayton1985} and shown to be effective and robust structures for accelerating electrons to ultra-relativistic energies by three research groups in 2004 \cite{Mangles2004, Geddes2004, Faure2004} and later for positrons by \cite{Corde2015}.

Accordingly, a natural application of our work—regarding plasma wakes driven by photon bursts via the Compton force—is the acceleration of electrons. At first glance, one might assume that insights from laser wakefield acceleration \cite{Esarey1996, Esarey2009} would suffice to conclude about leptons acceleration through Compton scattering. However, there are notable differences that merit emphasis. In this section, we will review the primary physical mechanisms known to limit electron acceleration—namely, wake phase velocity, dephasing, driver diffraction, and driver depletion. For each mechanism, we will provide a quantitative analysis and compare the findings to the established results in the laser-driven wakefield context.

\subsection{Collimated photon driver}
Previous work by the authors \cite{DelGaudio_PRL_2020} demonstrated that a radiation burst consisting of photons with wavelengths much shorter than the inter-particle distance of plasma electrons does not experience light dispersion, as the concept of a dielectric medium does not apply here. In fact, these photons do not perceive the plasma as a macroscopic, collective medium but rather as a rarefied gas. They propagate at the speed of light between successive Compton scattering events, which gradually deplete the photon population within the burst. Consequently, a perfectly collimated photon burst travels at the speed of light through the plasma for a time $t \ll \nu_{\omega}^{-1}$. This behavior is fundamentally different from the propagation of a typical laser pulse, which travels at its group velocity. In an unmagnetized plasma, electromagnetic wave dispersion limits the Lorentz factor associated with the wave’s group velocity to $\gamma_g = (1 - \beta_g^2)^{-1/2} = \omega / \omega_p$, where $\beta_g c$ is the group velocity and $\omega$ is the wave frequency. It is well established that an electrostatic wakefield excited by a laser driver exhibits a phase velocity $\beta_\phi = \beta_g$. The phase velocity of the wake critically determines the maximum energy gain that an accelerated charged particle can achieve. During acceleration, a particle can reach velocities exceeding the wake phase velocity and consequently outrun the plasma wave. When this occurs, acceleration ceases, and the maximum attainable particle energy is limited by dephasing. The phenomena of trapping, acceleration, and dephasing of charged particles have been extensively studied in one dimension \cite{Esarey_PoP_1995, Esarey1996, Esarey2009}. In both linear and nonlinear regimes, the maximum Lorentz factor of an accelerated charge scales as $\gamma_{\mathrm{max}} \propto \gamma_{\phi}^2$.

\begin{figure}
\includegraphics[width=0.8\linewidth, center]{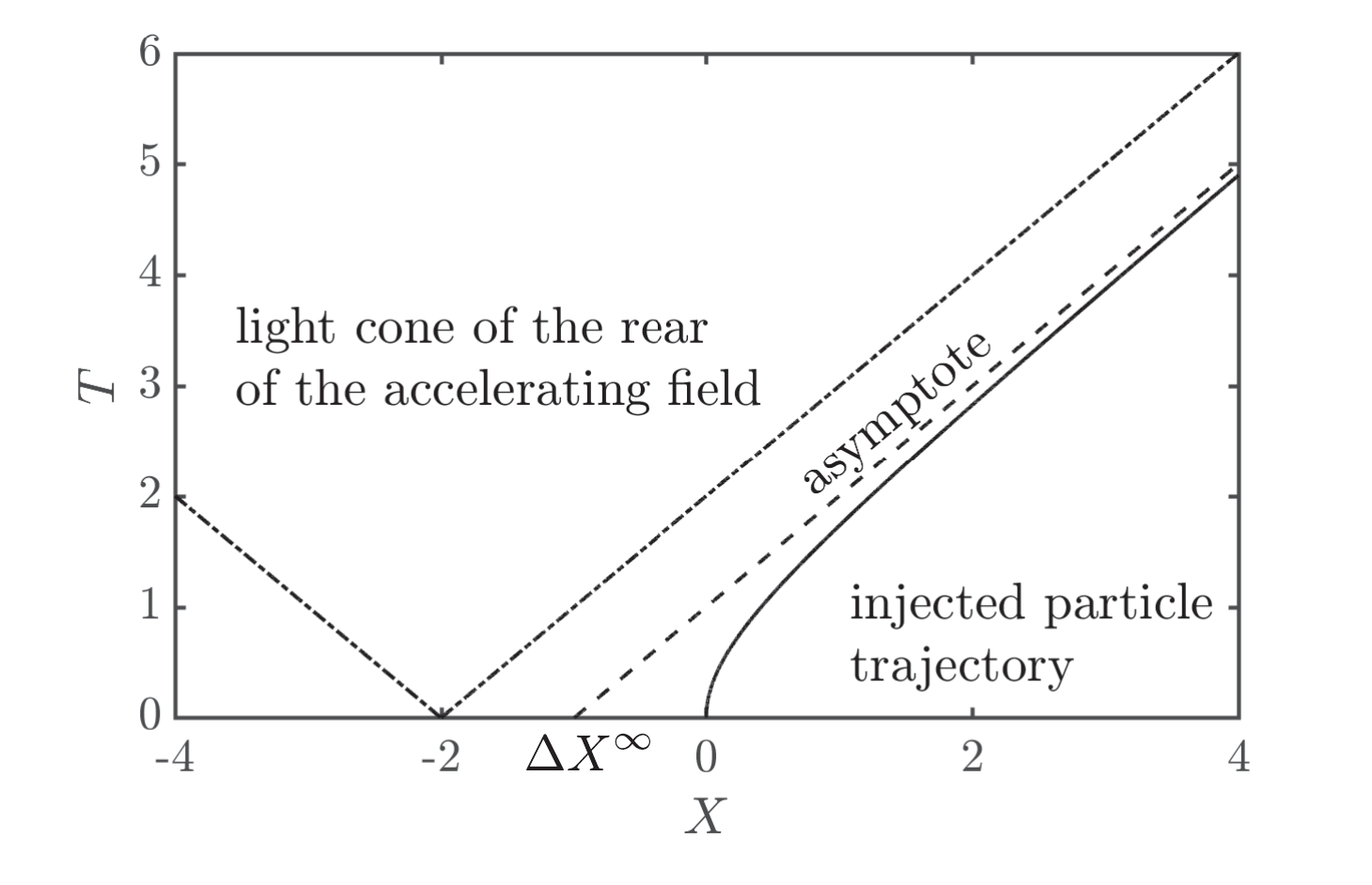}
\captionsetup{justification=raggedright}
\caption{Example of phase-locking condition. The particle trajectory (in solid line) converges to its asymptotic (in dashed line) $T=X+\Delta X^{\infty}$ and will never enter the light cone of the rear of the accelerating zone (dot-dashed), set at $X=-2$ initially.}
\label{fig: trapping}
\end{figure}

To examine the acceleration of electrons within the wake generated by a collimated photon burst, we initially neglect the depletion of the driver. The primary objective is to analyze the trapping conditions when the phase velocity equals the speed of light. In this scenario, an electron captured at the front of the accelerating region of the plasma wave cannot overtake the wave front at any subsequent time. Consequently, the electron will continuously accumulate a phase difference relative to the wake and eventually slip backwards, never outrunning the plasma wave. For simplicity, in the following model, the wake is assumed to have a square-shaped profile. Formally, the wakefield can be expressed as:
\begin{equation}
E(x,t) = E_x\sign \Big[\sin\big(k_p(x-ct)\big)\Big].
\end{equation}
A more rigorous calculation can be carried out with a sine function, but it turns out to be cumbersome without changing the final result. The accelerating field $E_x$ is thus constant on half a wavelength, and as long as the electron does not slip back. The motion $X(T) = e|E_x| x(t)/mc^2$ of an electron injected at the front $X_f$ at time $T = e|E_x| t/mc=0$ with initial momentum $p_0=mc\sqrt{\gamma_0^2-1}$ can be derived analytically.
The momentum evolves as $p= e|E_x|t + p_0$ and the energy as $\gamma = \sqrt{1+\left(T+p_0/mc\right)^2}$. Integrating the equation of motion leads to
\begin{eqnarray}
\frac{dX}{dT} &=& \frac{T+p_0/mc}{\sqrt{1+\left(T+p_0/mc\right)^2}}\\
X &=& \sqrt{1+\left(T+p_0/mc\right)^2}-\gamma_0.
\end{eqnarray}
The distance of the electron to the front of the acceleration region, which moves at the speed of light $X_f(T)=T$, is $\Delta X(T)= X_f(T)-X(T)$, or
\begin{equation}
\Delta X= T +\gamma_0 - \sqrt{1+\left(T+p_0/mc\right)^2}.
\end{equation}
The acceleration persists until the phase slip distance $l_\phi = mc^2 \Delta X/e|E_x|$ exceeds the length of the acceleration region, which we take to be half the plasma wavelength, i.e., $l_\phi > \lambda_p/2$. Furthermore, over sufficiently long times, the distance $\Delta X$ becomes effectively time-independent.
\begin{equation}
\label{eq:dxinfty}
\Delta X^{\infty}=\lim_{T\rightarrow\infty} \Delta X=\gamma_0-p_0/mc.
\end{equation}
This situation echoes the longitudinal invariance of an electron in a plane electromagnetic wave. If the asymptotic phase slip distance $l_{\phi}^{\infty} = mc^2\Delta X^{\infty}/e|E_x|$ remains shorter than the length of the acceleration region ($\lambda_p/2$), the electron will become phase-locked with the wake. This implies that the worldline of the electron always remains outside the light cone of the rear boundary of the accelerating zone, as illustrated in Fig.~\ref{fig: trapping}. The phase-locking condition, expressed in terms of the electric field of the wave, is given by:
\begin{equation}
\label{eq:Extrap}
\frac{|E_x|}{E_0} \gtrsim \frac{1}{\pi}\left(\gamma_0-\frac{p_0}{mc}\right)
\end{equation}
For an electron initially at rest, it reads $|E_x| \gtrsim E_0/\pi$, which is rather constraining since the field should be on the order of the cold wave-breaking limit. The condition for trapping and phase locking is considerably lower for an ultra-relativistic injected electron $\gamma_0 \gg 1$, which amounts to $E_x > E_0/(2\pi \gamma_0)$. We point out that these results are only valid for the special case $\beta_{\phi} = 1$. For LWFA, $\beta_{\phi} = \beta_g < 1$, although $\beta_g$ approaches unity for low plasma density or high frequency. 

\begin{figure}
\includegraphics[width=0.8\linewidth, center]{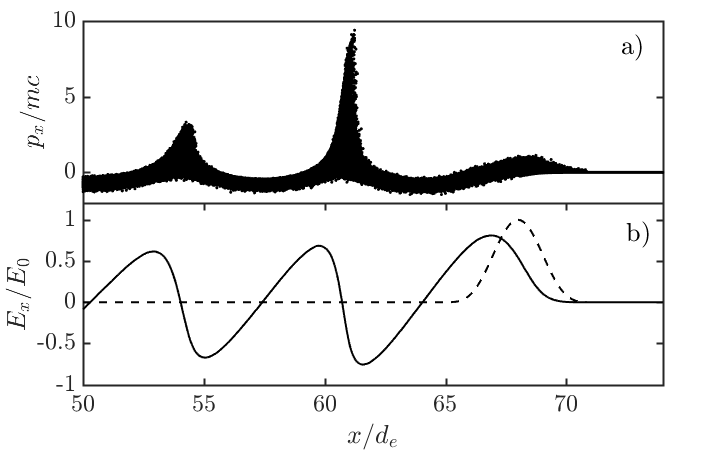}
\captionsetup{justification=raggedright}
\caption{One-dimensional simulation of electron acceleration in a wake driven by Compton scattering. A resonant burst of energy density $\mathcal{E}=eE_0/\sigma_T$ with 50 keV photons propagates in a plasma of density $n_p=10^{18}~\mathrm{cm^{-3}}$: a) Longitudinal momentum of the electrons $p_x$, b) Longitudinal wake field $E_x$ (solid line) and photon driver density in arbitrary units (dashed line).}
\label{fig: onset}
\end{figure}
We validated the electron acceleration process through Particle-In-Cell (PIC) simulations. In these simulations, the photon burst consists of $5\times 10^5$ macro-photons, each with an energy of 50 keV, corresponding to the normalized amplitude $\mathcal{E}_0=eE_0/\sigma_T$. The photons propagate along the $x$ axis and interact with an initially cold plasma of density $n_p=10^{18}~\mathrm{cm^{-3}}$, represented by 32 macro electrons per simulation cell. The simulations employ a moving window of length $24k_p^{-1}$, discretized into 24000 grid cells. The time step $\Delta t$ is set to $(1 - 10^{-6}) \Delta x$ to minimize the numerical dispersion of the wake, thus ensuring Courant stability. Figure \ref{fig: onset} displays the early stage of the acceleration process after a propagation distance of $15 k_p^{-1}$. Panel (a) shows the longitudinal momentum distribution of electrons, revealing characteristic signatures of trapping and acceleration near the rear of the wake. Panel (b) depicts the corresponding electrostatic field: the solid line represents the field behind the driver, shown by the dashed line.
In this simulation, the measured wakefield amplitude is $E_{\mathrm{max}}=0.74 E_0$, which aligns well with Eq. (\ref{eq:Emaxrel}) that predicts $E_{\mathrm{max}}\simeq0.78 E_0$. Since the cold wave breaking field is nearly reached, electrons can be trapped in the wake from rest, consistent with Eq. (\ref{eq:Extrap}). A detailed examination of the simulation data indicates that the trapped particles do not precisely occupy the region of maximum field. Instead, the maximum accelerating field experienced by the trapped electrons is $E_{acc}=0.61 E_0$. Additionally, simulations incorporating a $10\%$ energy spread in the photon driver were conducted; these show no significant impact on wake excitation or electron acceleration, as expected based on \cite{DelGaudio_PRL_2020}.

The prospect of electrons remaining phase-locked indefinitely implies that dephasing may not constrain the maximum energy in a Compton wakefield accelerator, unlike in laser or plasma-based schemes. However, this result is valid only for an idealized, non-evolving driver. For any realistic driver, multiple factors contribute to limiting the maximum achievable energy of an accelerated electron. In general, the maximum Lorentz factor can be expressed as: 
\begin{equation}\label{eq: gmax}
\gamma_{\mathrm{max}}\simeq \frac{E_x}{E_0}\frac{l_a}{d_e}
\end{equation}
where $l_a$ is the length over which the electron can be accelerated.
In the following sections, we examine the effects of driver depletion and diffraction.

\subsection{Driver depletion}
As the burst generates the plasma wake, the finite energy content of the driver is gradually depleted, which in turn impacts the sustainability of the plasma wake. In laser wakefield acceleration (LWFA), it is well established that the laser pulse depletes as it drives the plasma wave. The pump depletion length $L_{pd}$ can be estimated by equating the initial laser energy to the energy transferred to the wakefield as shown by \cite{Esarey1996, Shadwick2009, Esarey2009}, which leads to $L_{dp} \simeq (\lambda_p^3/\lambda)f(a_0)$ with $f(a_0) = 2/a_0^2$ if $a_0 \ll 1$ or $f(a_0) = \sqrt{2}/\pi a_0$ if $a_0 \gg 1$. 

The photons that transfer momentum to the plasma electrons are deflected and eventually exit the driver. Similar to the laser case, significant electron acceleration can only occur within the characteristic depletion length of the driver. This length corresponds to the mean free path of a photon, given by $l_{dp} = (\sigma_T n_p)^{-1}$. Consequently, the maximum Lorentz factor attainable is:
\begin{equation}
\gamma_{\mathrm{max}}\sim \frac{E_x}{E_0}\frac{l_{dp}}{d_e}~. \label{eq: gmax depletion}
\end{equation}

\begin{figure}
\includegraphics[width=0.8\linewidth, center]{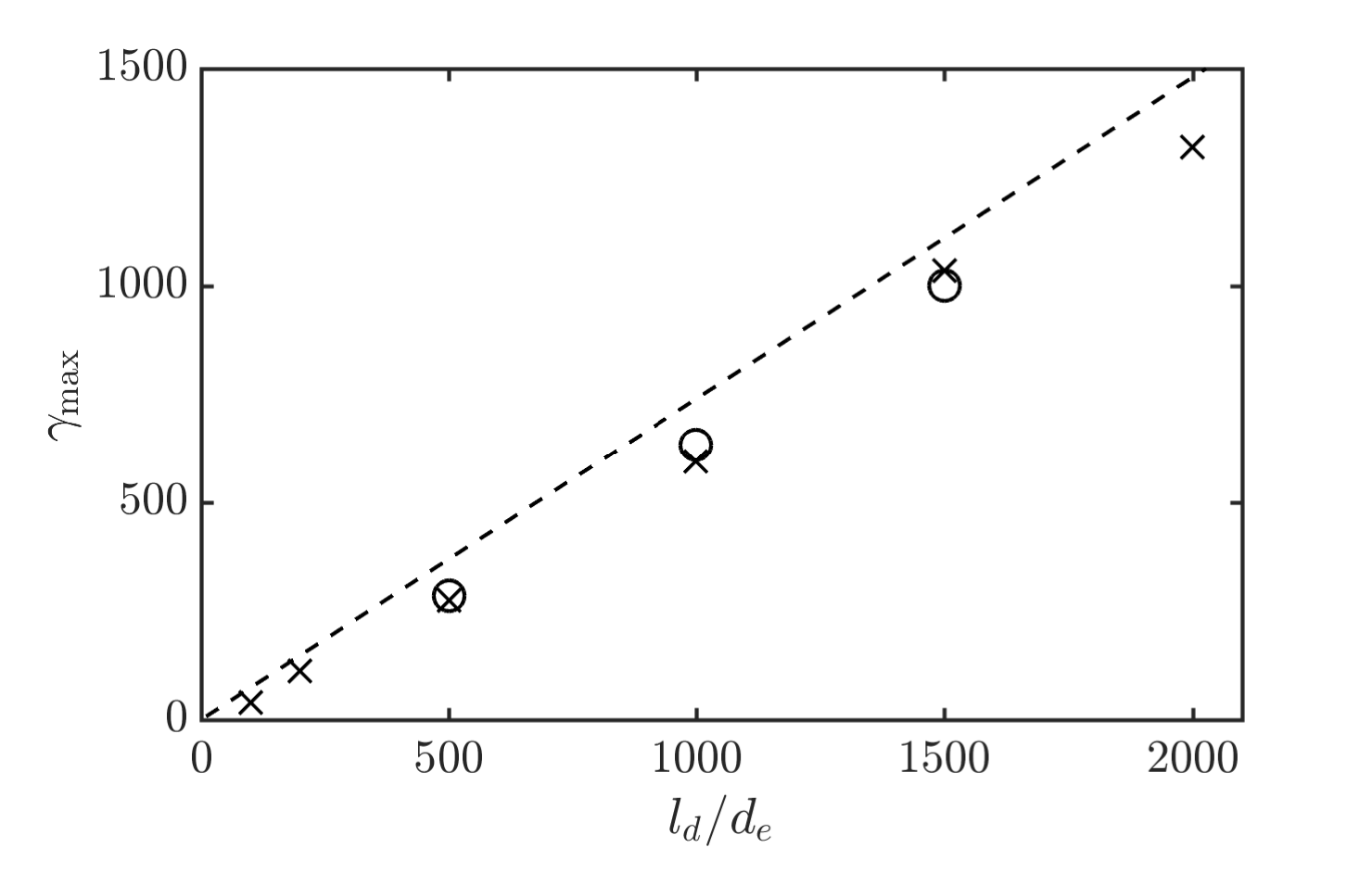}
\captionsetup{justification=raggedright}
\caption{Maximum Lorentz factor $\gamma_{\mathrm{max}}$ as a function of the driver depletion length $l_d$. Theory (dashed line) refers to the maximum field $E=0.74 E_0$. Simulations $(\times)$ with parameters: $\mathcal{E}=eE_0/\sigma_T$, 50 keV photons, and $n_p=10^{18}~\mathrm{cm^{-3}}$. The depletion length has been artificially reduced (with a factor $M=10^6$) to observe the saturation of the acceleration. The introduction of a $10\%$ energy spread in the photon driver ($\circ$) shows no influence on the acceleration process.}
\label{fig: depletion}
\end{figure}

\begin{figure}
\includegraphics[width=0.8\linewidth, center]{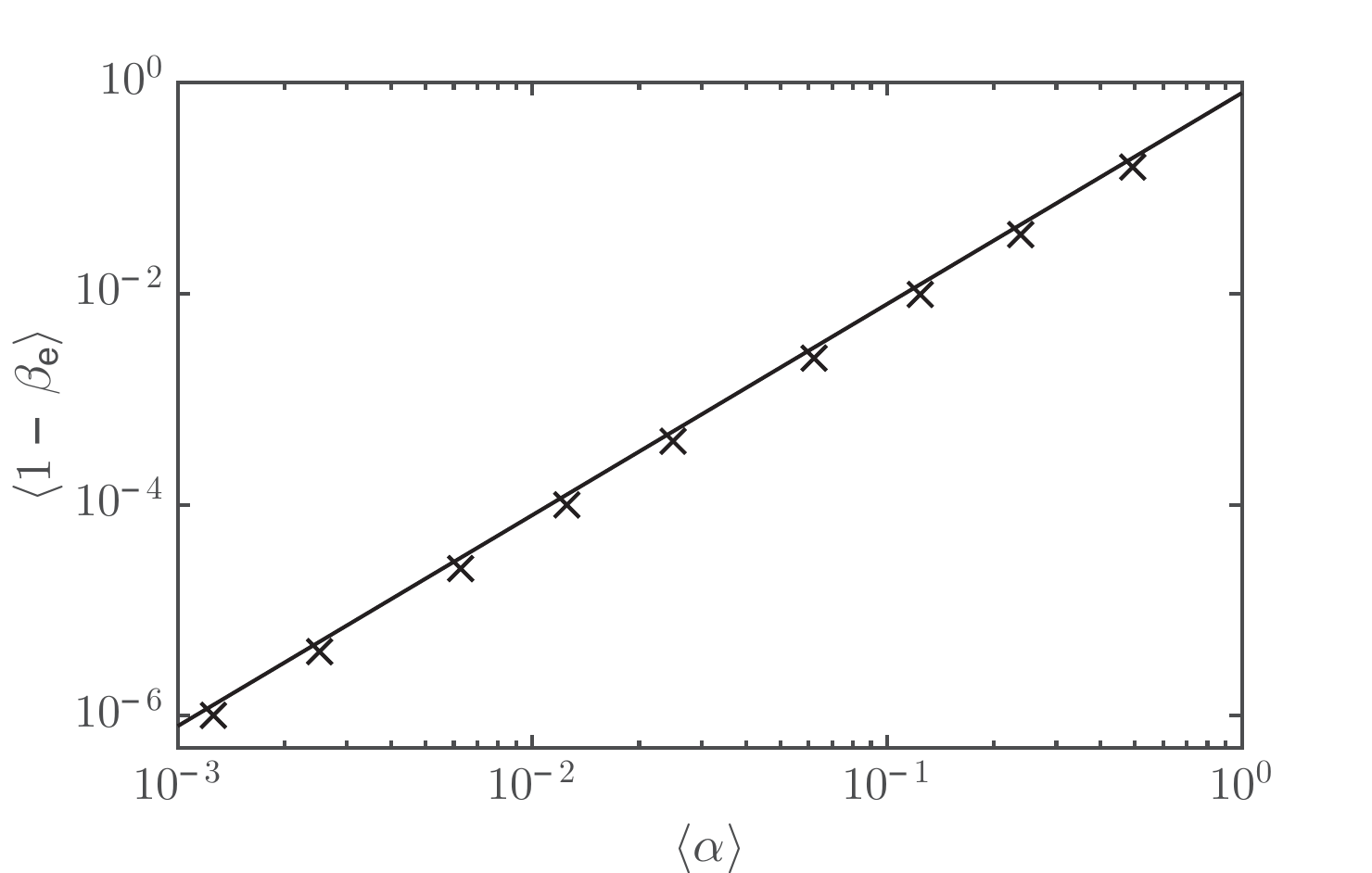}
\captionsetup{justification=raggedright}
\caption{Ensemble longitudinal velocity of the photon burst as a function of the divergence angle $\langle\alpha\rangle$. Equation (\ref{eq: DivAlpha}) in solid line and simulations in $(\times)$.}
\label{fig: DivAlpha}
\end{figure}

For a dense plasma of density $n_p=10^{18}~\mathrm{cm^{-3}}$ the electron inertial length is $d_e\simeq 5.3~\mu\mathrm{m}$ and the depletion length would be $l_{dp}\simeq 15.4~\mathrm{km}$. Conducting kinetic simulations across this vast scale separation is computationally unfeasible. To circumvent this limitation, we artificially increase the scattering cross-section by a factor $M=10^6$, such that $\sigma_T \rightarrow M \sigma_T$. This numerical rescaling enables us to observe the saturation of acceleration within a feasible simulation timeframe. Figure \ref{fig: depletion} illustrates the saturation of the maximum Lorentz factor $\gamma_{\mathrm{max}}$ measured after a propagation distance of approximately $l\simeq 2l_{dp}/d_e$. The acceleration ceases when the electric field diminishes to a level where the accelerated particles can no longer stay phase-locked with the wake. Notably, the simulation results agree with our estimate, derived from Eq. (\ref{eq: gmax depletion}), within a relative error of about $10\%$.

\subsection{Non-collimated driver}
A spread in the transverse component of the photon distribution is another effect that limits the acceleration of electrons.  If the driver is not collimated, its ensemble velocity is lower than the velocity of light and corresponds to the average velocity along the propagation direction $\hat{\bm k}_0$. Assuming a Gaussian spread $\sigma$ in the transverse direction, the photon distribution function can be written as
\begin{eqnarray}
\label{eq:spread}
\mathcal{N}({\bm k}) = \frac{1}{2\pi\sigma^2}\exp\left(-\frac{{\bm k}^2_{\perp}}{2\sigma^2}\right)\delta({\bm k}_{\parallel}-{\bm k_0}).
\end{eqnarray}
We define
\begin{eqnarray}
\langle \sin\alpha \rangle &=& \int~d{\bm k} \frac{k_{\perp}}{k_{\parallel}}\mathcal{N}({\bm k}) \\
\nonumber
&=&\sqrt{\frac{\pi}{2}}\frac{\sigma}{k_0}.
\end{eqnarray}
For $\sigma/k_0 \ll 1$, the average angle is $\langle \alpha \rangle \simeq \sqrt{\pi/2}(\sigma/k_0)$.
The ensemble velocity along ${\bm k}_0$ is related to $\alpha$ via $\langle\beta_e\rangle = \langle \cos\alpha\rangle$ which gives for $\alpha \ll 1$
\begin{eqnarray}
\label{eq: DivAlpha}
\langle\beta_e\rangle \simeq  1-\frac{1}{2}\langle\alpha^2\rangle = 1-\frac{2}{\pi}\langle\alpha\rangle^2.
\end{eqnarray}
The simulations presented in this section are performed with a moving window $36k_p^{-1}$ long, with resolution $\Delta x = 0.001~d_e$, and time step $\Delta t = 0.00099~\omega_p^{-1}$.
The number of particles per cell is $30$. The driver is composed of photons of energy $\hbar\omega=0.01~mc^2$ with $\mathcal{E}_0=eE_0/\sigma_T$ and the distribution is initialized according to Eq.(\ref{eq:spread}). The reference plasma density is always taken to be $n_p=1~\mathrm{cm^{-3}}$.

Figure \ref{fig: DivAlpha} illustrates the ensemble longitudinal velocity of the photon burst, which aligns with the predictions of Eq. (\ref{eq: DivAlpha}). As the driver’s energy is transferred to the plasma to excite the wake, the phase velocity of the wake $\beta_\phi$ is inherently equal to the ensemble velocity of the photon burst, i.e., $\beta_\phi = \beta_e$. In this scenario, electron acceleration would be limited by dephasing, analogous to the case in laser wakefield acceleration (LWFA). The dephasing length for linear and mildly nonlinear wakes is approximately $l_{d} \simeq \gamma_{\phi}^2\lambda_p$ \citep{Tajima_PRL_1979, Esarey_PoP_1995}, where $\gamma_\phi$ is the Lorentz factor associated with the wake phase velocity. Using Eq.(\ref{eq: DivAlpha}) lead to $\gamma_{\phi}^2=(\pi/4)/\langle\alpha\rangle^2$, which gives
\begin{equation} \label{eq: dephasing}
l_{d}\simeq\frac{\pi}{4}\frac{\lambda_p}{\langle\alpha\rangle^2}
\end{equation}
We emphasize that dephasing occurs systematically in laser plasma accelerators, even with non-evolving pulses. In contrast, for Compton-driven wakes, dephasing effects are only significant when the driver deviates from ideal conditions.
\begin{figure}
\includegraphics[width=0.8\linewidth, center]{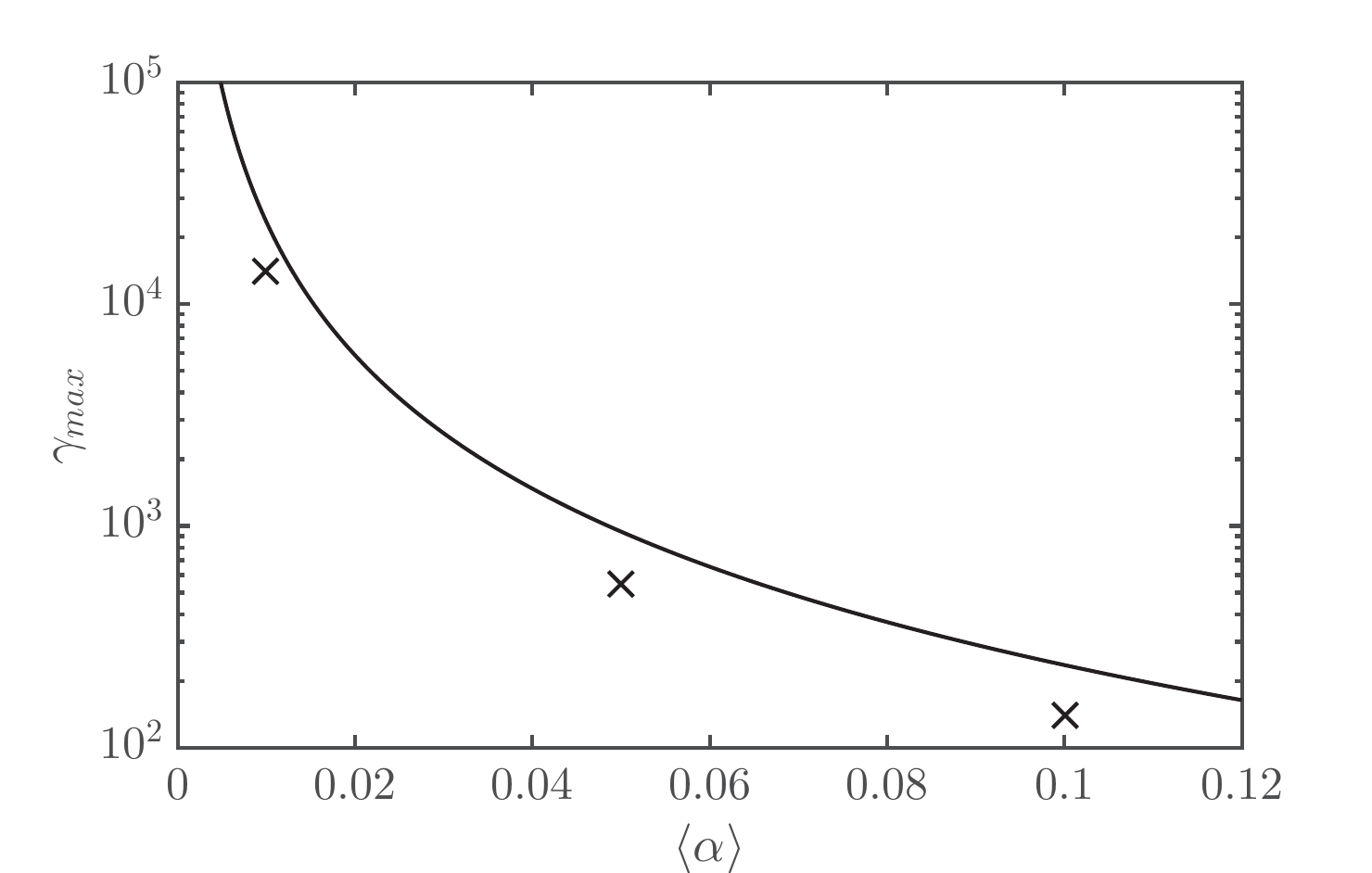}
\caption{Maximum Lorentz factor as a function of $\langle\alpha\rangle$ given by dephasing.
Simulations in $(\times)$ and Eq. (\ref{eq: gmax}) in solid line, where $l_a=l_{d}$ and $E_x = 0.61E_0$ corresponding to a resonant driver energy density of $\mathcal{E}_0=eE_0/\sigma_T$.}
\label{fig: max_g_dephasing}
\end{figure}

\section{Two-dimensional effects}
\label{sec:2d}

\begin{figure}
    \subfigure[\ Plasma electron density.]{
    \includegraphics[width=0.5\linewidth]{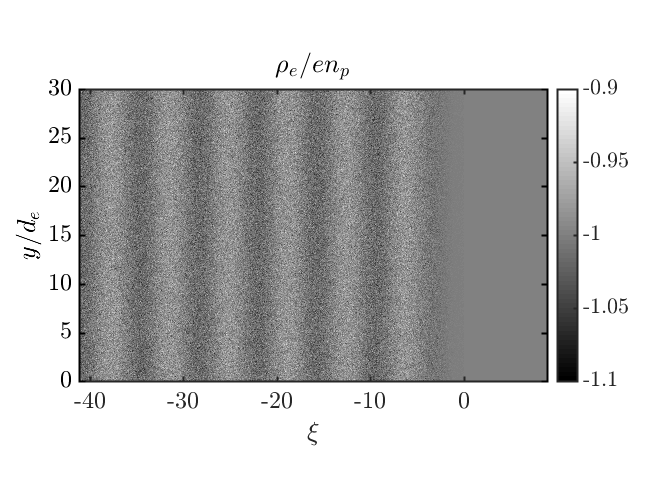}}
    \subfigure[\ Longitudinal electric field.]{
    \includegraphics[width=0.5\linewidth]{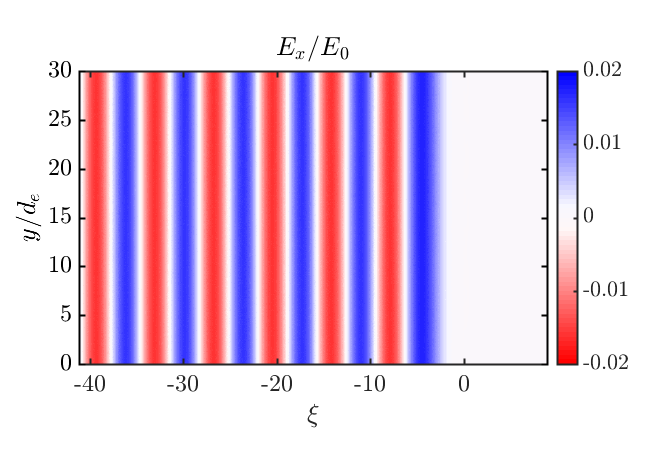}}\\
    \subfigure[\ Transverse focusing field on a relativistic charge.]{
    \includegraphics[width=0.5\linewidth, center]{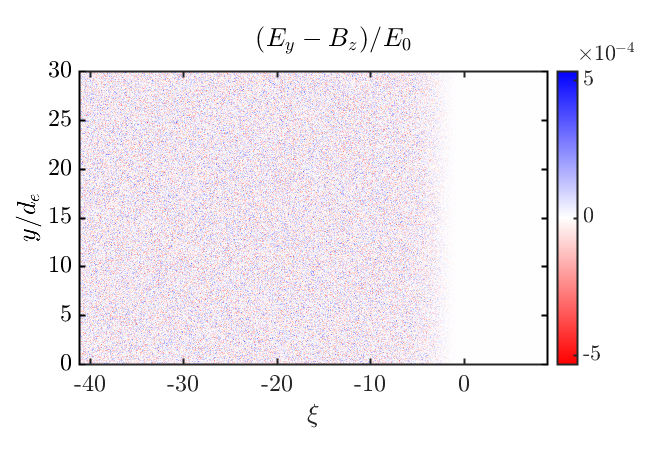}}
    \captionsetup{justification=raggedright}
    \caption{2D simulation of Compton wakefield linear regime ($\mathcal{E}_0=0.01~eE_0/\sigma_T$) with driver transverse size $L_y\gg\lambda_p$.}
    \label{fig: LI}
\end{figure}
\begin{figure}
    \subfigure[\ Plasma electron density.]{
    \includegraphics[width=0.5\linewidth]{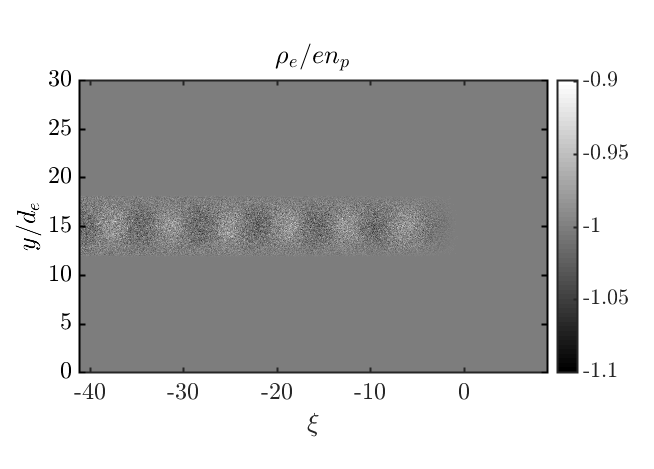}}
    \subfigure[\ Longitudinal electric field.]{
    \includegraphics[width=0.5\linewidth]{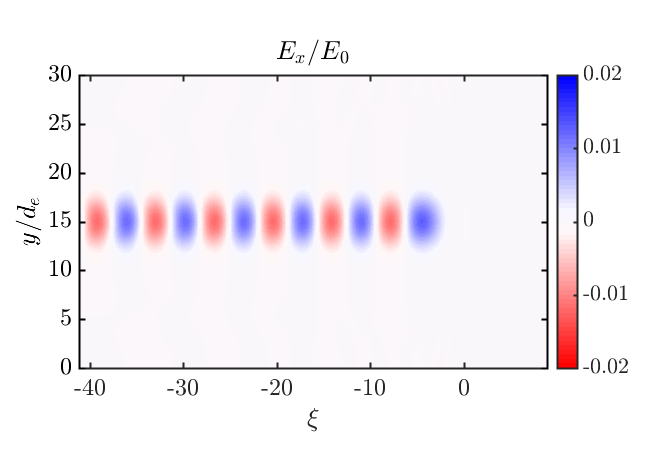}}\\
    \subfigure[\ Transverse focusing field on a relativistic charge.]{
    \includegraphics[width=0.5\linewidth, center]{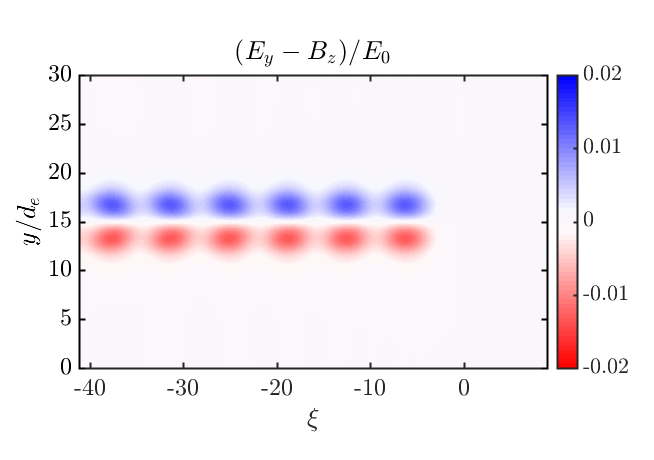}}
    \captionsetup{justification=raggedright}
    \caption{2D simulation of Compton wakefield linear regime ($\mathcal{E}_0=0.01~eE_0/\sigma_T$) with driver transverse size $L_y\simeq\lambda_p$.}
    \label{fig: LR}
\end{figure}
The one-dimensional theory and simulations of Compton wakes assume that the photon driver has an arbitrarily large transverse extent. However, the longitudinal properties of these wakes should remain valid as long as the driver’s width exceeds its length. It is well known from laser wakefield theory that finite beam width introduces focusing fields in the wake, and in the limit of the blowout regime, the wake adopts a bubble-shaped structure as first observed by \cite{Pukhov2002}, and further investigated in depth theoretically \cite{WLU2006}, and numerically \cite{WLu2007}. This regime is considered optimal for electron acceleration but requires precise tuning of the laser pulse profile with the plasma density. Consequently, such conditions are unlikely to be readily achievable outside controlled laboratory settings. A similar reasoning could be applied to a photon driver. Nonetheless, for the sake of curiosity, it is legitimate to question whether the properties of two-dimensional wakes generated by laser pulses differ from those produced by photon drivers. To explore this, we numerically investigate two-dimensional wake structures generated by various photon drivers, varying both the amplitude and the transverse width, in both the linear and nonlinear regimes. For simplicity, we restrict our study to collimated photon drivers. In the case of laser-driven plasma wakes, the transverse fields can generally be estimated from the longitudinal fields using the Panofsky-Wenzel theorem \cite{Panofsky1956}, assuming electrons are primarily accelerated along the longitudinal direction and are sufficiently relativistic. In two-dimensional Cartesian geometry, and in the limit where $\beta_x \to 1$, the theorem states that: $\partial_{\xi}(E_y-B_z) \simeq \partial_y E_x$. As we will see, the theorem does not hold for a Compton-driven wake.

\subsection{2D Linear regime}
The two-dimensional simulations are performed in a moving window $50d_e\times30d_e$ long and wide, respectively, with resolution $\Delta x = 0.01 d_e$, and time step $\Delta t = 0.007\omega_p^{-1}$. The number of particles in each cell is $256$. The driver is always resonant and composed of photons of energy in the range $\hbar\omega=0.001-0.01 mc^2$, and the reference plasma density is $n_p=1~\mathrm{cm^{-3}}$. All the figures show: a) the plasma density, b) the longitudinal electric field, and c) the total transverse field.

The simulations with an infinite width are performed with periodic boundary conditions. Figure \ref{fig: LI} corresponds to a driver in the linear regime $\mathcal{E}_0=0.01~eE_0/\sigma_T$. In this regime, the amplitude of the wake is $E_x = (\pi/2)\sigma_T\mathcal{E}_0/e$, which gives $E_x/E_0 = (\pi/2)\times 10^{-2}$ for the parameter of the simulation, which is in agreement with Fig.\ref{fig: LI}-b). The transverse fields shown in  Fig.\ref{fig: LI}-c) vanish as expected in the linear regime for an infinitely wide driver since $\partial_y E_x = 0$.

For a round photon driver that possesses in both longitudinal and transverse directions an identical density profile 
\begin{equation}
\mathcal{E}(x,y,t)= \mathcal{E}_0\sin^2(\xi/2)\cos^2(\pi y/L_y),
\end{equation}
with $\xi = k_p(x-ct) \in [-2\pi,0]$, and $y \in [-L_y/2,L_y/2]$, and $L_y = 2\pi/k_p$. Figure \ref{fig: LR} shows the key properties of the wake. In the linear theory of LWFA ($a_0 \ll 1$), \cite{Esarey2009} summarizes the following scalings: $E_x \sim E_y \propto a_0^2$, and $B_z \propto a_0^4$  and one could expect $E_x \sim E_y \propto \mathcal{E}_0$, and $B_z \propto \mathcal{E}_0^2$. The longitudinal electric field retains the one-dimensional form within the driver envelope, approximately, $E_x \simeq (\sigma_T \mathcal{E}_0/e)\sin(\xi)\cos^2(\pi y/L_y)$. 
Although $E_y$ can be estimated by the transverse derivative of $E_x$, i.e, $E_y \simeq k_p^{-1}\int d_{\xi} \partial_y E_x$, the magnetic field exhibits a distinct structure. Part of the longitudinal current produced when photons scatter off fresh plasma is redirected transversely owing to the stochastic character of the Compton force, producing a DC current component resembling the return currents on either side of the driver. The AC part of the current obeys $\partial_{\xi}E_x \simeq 4\pi J_{AC}/\omega_p$, and the DC part $\langle J_x \rangle_x = J_{DC} = \partial_y B_z/(4\pi c)$. Consequently, $B_z$ is bipolar with a magnitude comparable to $E_y$, which explains the observed $B_z$ morphology. The DC contribution of $B_z$ produces an offset such that $\langle E_y-B_z\rangle_{x} \neq 0$. For wider beams, the average DC current vanishes, and the imbalance is not observed.

\subsection{2D Non-linear regime}

\begin{figure}
    \subfigure[\ Plasma electron density.]{
    \includegraphics[width=0.5\linewidth]{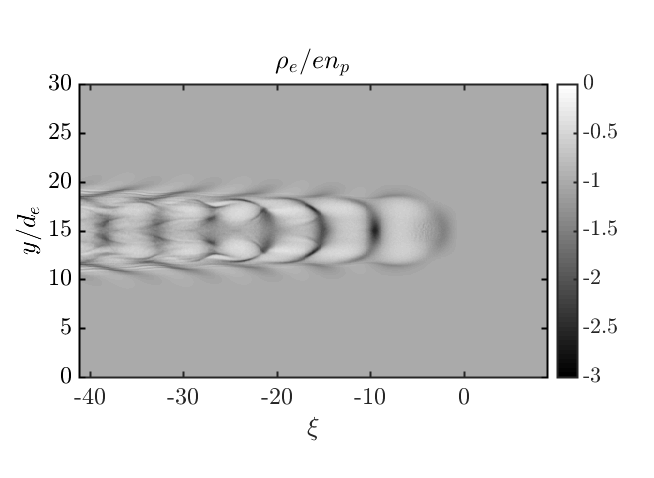}}
    \subfigure[\ Longitudinal electric field.]{
    \includegraphics[width=0.5\linewidth]{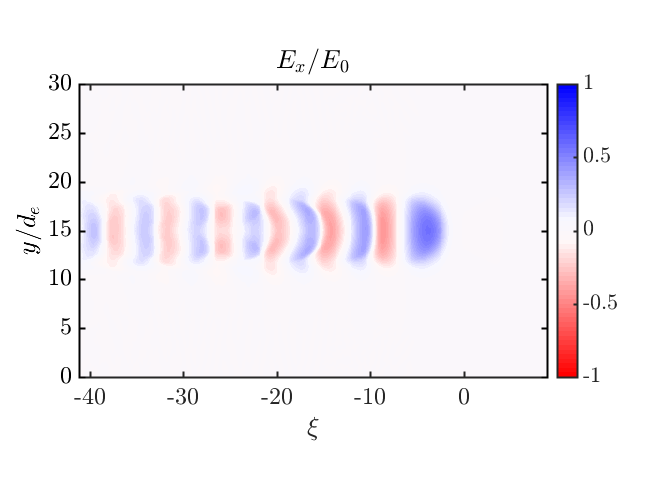}}
    \subfigure[\ Transverse focusing field on a relativistic charge.]{
    \includegraphics[width=0.5\linewidth, center]{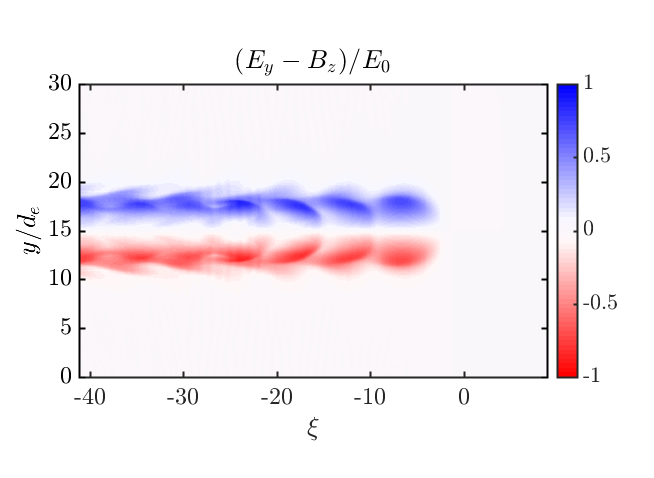}}
    \captionsetup{justification=raggedright}
    \caption{2D simulation of Compton wakefield nonlinear regime ($\mathcal{E}_0=eE_0/\sigma_T$) with driver transverse size $L_y\simeq\lambda_p$.}
     \label{fig: NR}
\end{figure}

As stated earlier in the article, nonlinear wakes are excited when $\mathcal{E}_0 \sim eE_0/\sigma_T$. Using the same parameter set as before, we excited a nonlinear wake with a finite-width photon driver shown in Figure \ref{fig: NR}. The previously proposed explanation for the emergence of a DC magnetic component remains valid, and the net transverse field acting on a relativistic electron is therefore always focusing. The characteristic bubble morphology familiar from laser wakefield accelerators is only partially reproduced. Because the Compton driving force acts predominantly along the longitudinal coordinate $\xi$, whereas a laser ponderomotive force displaces the electron fluid comparably in longitudinal and transverse directions, the resulting electron cavity is more elongated.

\section{Discussion and conclusions}
\label{sec:conclusions}
\subsection{Summary}
In this work, we have explored the generation of plasma wakes by non-ponderomotive drivers, specifically photon bursts where the dominant interaction is Compton scattering. This regime is realized when the photon wavelength is much smaller than the inter-particle distance but still exceeds the Compton wavelength $\lambda_C$. When $\lambda < \lambda_C$, the momentum transfer becomes significant, allowing the electrons to be driven to relativistic velocities and effectively stream with the photons. 
We reviewed the linear theory for photon-burst-driven wakes and extended it to non-symmetrical drivers. Our findings indicate that plasma waves can attain amplitudes comparable to those driven resonantly, with $E \sim \sigma_T \mathcal{E}/e$, provided the photon burst’s onset and decay occur over lengths on the order of the resonant length. As detailed in our prior work \cite{DelGaudio_PRL_2020}, well-defined linear wakes are observable when $\nu_c > \omega_p$. The criteria for nonlinear Langmuir wake formation mirror those established for laser or particle drivers, notably when $\mathcal{E}_0 \gtrsim eE_0/\sigma_T$. Interestingly, even in extreme conditions where $\mathcal{E}_0 \gg eE_0/\sigma_T$, the wake amplitude remains proportional to the energy density, as for the linear theory.
Acceleration in Compton wakes present several important characteristics that emerge: (i) for a hypothetical perfectly collimated driver \footnote{A perfectly collimated photon driver does not exist in Nature. It is a theoretical object to study the limit of Compton wake acceleration}. The wake propagates at exactly the speed of light; (ii) electrons can phase-lock within the wake, though the photon driver’s depletion length ultimately limits their acceleration; (iii) non-collimated drivers introduce an effective subluminal phase velocity, with acceleration constrained by the corresponding dephasing length.
Our 2D simulations revealed additional nuanced behaviors: although the longitudinal component of the wake aligns well with 1D predictions, the transverse fields differ markedly from those observed in conventional laser wakefields. This discrepancy stems from the plasma response to the driver, which induces a longitudinal current along the axis and results in a DC magnetic field buildup behind the wake. Consequently, the transverse fields are consistently focused at all positions within the wake, highlighting a unique feature of this regime.

\subsection{Compton wakes in the laboratory and in astrophysics}
The possible observation of Compton wakes in the laboratory requires high frequency photons and a driver with a length the order of the plasma wavelength. The maximum photon energy density for the X-ray Free Electron Laser is typically on the order of $10^{12}~\mathrm{erg/cm^3}$. Eq.(\ref{eq:Ene_den_nl}) reveals that the nonlinear regime could only be accessible for extremely low density plasmas $n_p \sim \mathrm{cm^{-3}}$. The plasma would have to be several kilometers long, which makes this nonlinear behavior out of reach in the laboratory with current technology. 
As discussed in our previous work, linear Compton wakes may be observed  in the future for a plasma density $n_p \sim \mathrm{10^{17}/cm^{3}}$ with a non-resonant driver of length $L / \lambda_p=10$ (duration of 40 fs), and 100 eV X-rays. The required energy density would be $\mathcal{E} \sim 10^{16}~\mathrm{erg/cm^3}$, and the total energy of the driver would be on a few Joules, which is still orders of magnitude above the energy on a X-ray Free Electron Laser. 

Whereas low-density and very long plasmas are hard to produce in the laboratory, they are naturally present in the interstellar medium. We should therefore determine whether the observed extreme luminosities could potentially correspond to a photon energy density capable of driving nonlinear Compton wakes. Considering a spherical object of radius $r_0$ and surface $4\pi r_0^2$, the luminosity and the radiation energy density are related by 
\begin{equation}
    \mathcal{E}= \frac{\mathcal{L}}{4\pi r_0^2c} \sim 10^{14}\frac{\mathcal{L}[\mathcal{L}_{edd}]}{r_0^2[10^6 \mathrm{cm}]} \mathrm{erg/cm^3}.
\end{equation}
Comparing this expression with Eq.(\ref{eq:Ene_den_nl}) indicates that radiation from extremely luminous compact objects $\mathcal{L} \gg \mathcal{L}_{edd} \sim 10^{38} \mathrm{erg/s}$, propagating into low–density plasmas, can drive the development of nonlinear plasma wakes. The acceleration efficiency follows from the scaling derived in Sec.{\ref{sec:acceleration}}. Equation {\ref{eq: gmax depletion}} with $E_x \sim E_0$ can be expressed as $\gamma_{\mathrm{max}} \sim (n_pr_e^3)^{-1/2}.$, which applied to a plasma of $n_p \sim \mathrm{cm^{-3}}$ yields unphysically large Lorentz factors of $10^{21}$. More realistically, the photon burst should have an energy density that decreases during the propagation due to geometric dilution of the photon surface, $\mathcal{E} \sim \mathcal{E}_0r_0/(r_0+ct)^2$, which reduces the attainable Lorentz factor to
\begin{equation}
\gamma_{\mathrm{max}} \sim \frac{r_0}{d_e} = r_0[10^6 \mathrm{cm}]\sqrt{n_p[\mathrm{cm^{-3}}]}.
\end{equation}
Maximum acceleration efficiency is achieved for resonant drivers, implying that for very tenuous plasmas, the photon-burst duration must be shorter than a few milliseconds.

\section{Acknowledgments}
We acknowledge valuable discussions with Dr. Pablo Bilbao and Prof. Anatoly Spitkovsky. This work was supported by the European Research Council under Grant No. 695088 (ERC-2015-AdG) and by the Portuguese Foundation for Science and Technology (FCT) through grants No. PD/BD/114323/2016 within the framework of the Advanced Program in Plasma Science and Engineering (APPLAuSE, FCT Grant No. PD/00505/2012), and the project X-MASER - 2022.02230.PTDC. We also acknowledge PRACE and Euro HPC for providing access to the MareNostrum supercomputing resources in Spain. Computational simulations were conducted on the IST cluster in Portugal and the MareNostrum supercomputer in Spain.

\appendix

\section{Relativistic Compton force from the photon-electron collision operator}
\label{app: CMP_rel_force}
The electron distribution function $f(t,{\bm x},{\bm p})$ in the presence of collisions obeys the Boltzmann equation
\begin{equation}
    \frac{\partial f}{\partial t} + \frac{\bm p}{m}\cdot\nabla f + {\bm F}\cdot\nabla_{\bm p} f = \left(\frac{\partial f}{\partial t}\right)_{\mathrm{coll}}.
\end{equation}
The photon-electron collision operator, which accounts for Compton scattering, reads
\begin{equation}
\left. \frac{\partial f}{\partial t} \right\lvert_c = c\int d{\bm k}d\hat{\bm \Omega}_0'\frac{\omega}{\gamma\omega_0}\frac{d\sigma_0}{d\hat{\bm \Omega}_0'}\left(\mathcal{N}'f'-\mathcal{N}f\right)
\end{equation}
where $\mathcal{N}$ denotes the photon distribution function. In the case of small momentum exchange we can expand the distribution as $f'\simeq f+\delta{\bm p}\cdot\nabla f$, which gives
\begin{equation}
\nonumber
\left. \frac{\partial f}{\partial t} \right\lvert_c = c\int d{\bm k}d\hat{\bm \Omega}_0'\frac{\omega}{\gamma\omega_0}\frac{d\sigma_0}{d\hat{\bm \Omega}_0'}\left[\left(\mathcal{N}'-\mathcal{N}\right)f+\mathcal{N}'\delta{\bm p}\cdot\nabla f\right].
\end{equation}
If the driver is quasi-static, then $\mathcal{N}'\simeq\mathcal{N}$ and,
\begin{equation}
\left. \frac{\partial f}{\partial t} \right\lvert_c = c\int d{\bm k}d\hat{\bm \Omega}_0'\frac{\omega}{\gamma\omega_0}\frac{d\sigma_0}{d\hat{\bm \Omega}_0'}\mathcal{N}\delta{\bm p}\cdot\nabla f.
\end{equation}
The relativistic transformations of frequency and angle are given by
\begin{eqnarray}
\nonumber
\omega_0 &=& \gamma\omega(1-{\bm \beta}\cdot\hat{\bm \Omega}) \\
\nonumber
d\hat{\bm \Omega}_0'&=&\frac{d\hat{\bm \Omega}'}{\gamma^2(1-{\bm \beta}\cdot\hat{\bm \Omega}')^2},
\end{eqnarray}
and the small momentum exchange is
\begin{equation}
\delta{\bm p}\simeq\frac{\hbar\omega}{c}\left(\hat{\bm\Omega}-\hat{\bm\Omega}'\right).
\end{equation}
We arrive at the following expression for the operator
\begin{equation}
\nonumber
\left. \frac{\partial f}{\partial t} \right\lvert_c = \int \hbar\omega^3d\omega d\hat{\bm \Omega} d\hat{\bm \Omega}_0'(1-{\bm \beta}\cdot\hat{\bm \Omega})\frac{d\sigma_0}{d\hat{\bm \Omega}_0'}\mathcal{N}\left(\hat{\bm\Omega}-\hat{\bm\Omega}'\right)\cdot\nabla f.
\end{equation}
In the Thomson regime, where the energy of the photon in the rest frame of the electron is small $\hbar\omega_0\ll mc^2$ , the differential cross-section can be approximated as
\begin{equation}
\frac{d\sigma_0}{d\hat{\bm \Omega}_0'}\simeq \frac{r_e^2}{2}\left[1+\left(\hat{\bm\Omega}_0\cdot\hat{\bm\Omega}_0'\right)^2\right]
\end{equation}
If we consider a collimated photon distribution function along $\hat{\bm k}$, i.e., $\mathcal{N}=\mathcal{N}(\omega)\delta\left(\hat{\bm\Omega}-\hat{\bm k}\right)$, and  electrons initially moving along $\hat{\bm k}$ such that ${\bm \beta}\cdot\hat{\bm k}\simeq \beta$, then $\hat{\bm \Omega}_0=\hat{\bm \Omega}=\hat{\bm k}$. The total cross-section is the Thomson cross-section
\begin{equation}
\int d\hat{\bm \Omega}_0'  \frac{d\sigma_0}{d\hat{\bm \Omega}_0'} = -\sigma_T.
\end{equation}
The photon energy density is defined as
\begin{equation}
\int \hbar\omega^3d\omega \mathcal{N}(\omega) = \mathcal{E}.
\end{equation}
We finally obtain the photon-electron collision operator 
\begin{eqnarray}
\nonumber
\left. \frac{\partial f}{\partial t} \right\lvert_c &=& -\sigma_T\int d\hat{\bm \Omega} (1-{\bm \beta}\cdot\hat{\bm \Omega})\mathcal{E}\delta\left(\hat{\bm\Omega}-\hat{\bm k}\right)\hat{\bm\Omega}\cdot\nabla f \\
&=&-\sigma_T(1-{\bm \beta})\mathcal{E}\hat{\bf k}\cdot\nabla f,
\end{eqnarray}
where the term in front of $\nabla f$ should be interpreted as the relativistic version of the Compton force.

\bibliographystyle{jpp}
\bibliography{CMP_Nlin.bib}

\end{document}